\documentclass[12pt,a4paper]{article}
\usepackage{epsfig}
\title{On the DLCQ as a light-like limit \\
	in string theory}
\author{{\sc Shozo
Uehara}\footnote{e-mail: uehara@eken.phys.nagoya-u.ac.jp}{} ~and
{\sc Satoshi Yamada}\footnote{e-mail:
yamada@eken.phys.nagoya-u.ac.jp}\vspace{4mm}\\
{\sl Department of Physics, Nagoya University} \\
{\sl Chikusa-ku, Nagoya 464-8602, Japan}}
\date{}

\makeatletter

\@addtoreset{equation}{section}
\renewcommand{\thefigure}{\@arabic\c@figure}
\makeatother

\begin{document}
\maketitle
\vspace{-80mm}
\begin{flushright}
	DPNU-00-26\\
	July 2000
\end{flushright}
\vspace{50mm}
%%%%%%%%%%%%%%%%%%%%%%%%%%%%%%%%%%%%%%%%%%%%%%%%%%%%%%%%%%%%%%%%%
\begin{abstract}
We study the issue of defining the discrete light-cone quantization
(DLCQ) in perturbative string theory as a light-like limit.
While this limit is unproblematic at the classical level, it is
non-trivial at the quantum level due to the divergences by the
zero-mode loops. We reconsider this problem in bosonic string theory.
We construct the multi-loop scattering amplitudes in both open and
closed string theories by using the method of Kikkawa, Sakita and
Virasoro (KSV), and then we show that these scattering amplitudes are
perfectly well-defined in this limit.
We also discuss the vacuum amplitudes of the string theory.
They are, however, ill-defined in the light-like limit due to the
zero-mode loop divergences, and hence we want supersymmetry to cure
those pathological divergences even in string theory.
\end{abstract}
%%%%%%%%%%%%%%%%%%%%%%%%%%%%%%%%%%%%%%%%%%%%%%%%%%%%%%%%%%%%%%%%%%%
\section{Introduction}
%%%%%%%%%%%%%%%%%%%%%%%%%%%%%%%%%%%%%%%%%%%%%%%%%%%%%%%%%%%%%%%%%%%
Matrix theory \cite{BFSS} is a supersymmetric quantum mechanics
with matrix degrees of freedom.
This was proposed as a hamiltonian of M-theory in the infinite
momentum frame and it is expected to describe the fundamental degrees
of freedom in the large $N$ limit.
Furthermore Susskind proposed that the discrete light-cone
quantization (DLCQ) of M-theory is described by a finite $N$ Matrix
theory \cite{Sus}.

In DLCQ the compactified coordinate is the light-like coordinate
$x^-(\simeq x^-+2\pi R)$ and the corresponding momentum is quantized
as $p^+=N/R$.
To prove the Susskind's conjecture, Seiberg described a light-like
compactification as a limit of the compactification on a space-like
circle \cite{Sei,Sen}. Following ref.\cite{HP}, we will henceforth
refer to the Seiberg's limit as a light-like limit (L$^3$) to
distinguish it from the conventional DLCQ \cite{MY}. (We will also
refer to the conventional DLCQ as the direct DLCQ. )

Several people examined the L$^3$ of scattering amplitudes in field
and string theories compactified on an almost light-like circle
\cite{HP,HMV} or, Lorentz-equivalently, a vanishingly small
space-like circle with fixed $N$ \cite{Bil1,Bil2,Bil3,HMV}.
They found that the limit is, in general, complicated in field theory,
but it is well-defined at one-loop level in Type II string theory.
The problem in field theory is that when any external momenta in the
compact direction\footnote{Any external lines must have non-vanishing
momenta in the compact direction \cite{Bil3}.\label{foot1}} do not
flow through the loop, the zero-mode contribution in the sum over the
discrete loop momentum, which corresponds to the loop diagram of only
zero-mode propagators, is divergent as $\delta(0)$.
Hence we can call this divergence the zero-mode loop divergence.
  In Type II string theory this type of divergence does not occur at
the perturbative one-loop level.
  It was first pointed out that it would be due to the existence of
the winding modes \cite{Bil1}, and then it was pointed out that
it is because the closed string theory has only three-point
vertex \cite{Bil3}. Furthermore the existence of the light-like limit
in string theory to all orders in perturbation was discussed in
ref.\cite{Bil2}.
But it is not obvious whether the result of the one-loop scattering
amplitude can be applied to multi-loop scattering amplitudes without
keeping the integrals, or sums, over loop momenta.
To make this point clear is one of the purposes of this paper.

The vacuum energy in a theory which includes gravity is the
cosmological constant and, in string theory, we must also investigate
the vacuum amplitudes in the L$^3$. This is also one of the purposes
of this paper.
 
The plan of this paper is as follows. In section 2 we briefly discuss
the relevant kinematics. In section 3, we review the L$^3$ in field
theory and in section 4, we construct the bosonic string multi-loop
scattering amplitudes by the method of Kikkawa, Sakita and Virasoro
(KSV) \cite{KSV}. We show that these amplitudes are well-defined in
the L$^3$. In section 5 we investigate the vacuum amplitudes in the
L$^3$. Section 6 is devoted to our conclusion and discussions.
In appendices A and B, we compare $N$-point tree and one-loop
amplitudes by the KSV method with those of the operator formalism,
respectively. We clarify how the amplitudes include self-crossing
lines.
%%%%%%%%%%%%%%%%%%%%%%%%%%%%%%%%%%%%%%%%%%%%%%%%%%%%%%%%%%%%%%%%%%%%
\section{Kinematics}
%%%%%%%%%%%%%%%%%%%%%%%%%%%%%%%%%%%%%%%%%%%%%%%%%%%%%%%%%%%%%%%%%%%%
  We consider a 26-dimensional space-time coordinate system
($x^0 , x^1 ,x^i$), ($i=2,\cdots,25$). The space-like coordinate $x^1$
takes values on a circle of radius $R_s = \epsilon R$, where we have
introduced a parameter $\epsilon$ to consider $\epsilon\rightarrow 0$
limit with $R$ fixed, so that the corresponding momentum $p^1$ is
quantized, while the other coordinates $x^0, x^i$ are non-compact:

\begin{equation}
\left(\begin{array}{c}
	x^1\\ x^0
	\end{array}
	\right)
\simeq \left(\begin{array}{c}
	x^1\\ x^0
	\end{array}\right) + \left( \begin{array}{c}
				2\pi R_s\\ 0
				\end{array}\right),
   \hspace{3ex} x^i \simeq x^i, \hspace{3ex} p^1=\frac{N}{R_s}\,.
   \label{2.1}
\end{equation}

By the Lorentz boost with a boost parameter
$\beta = \frac{1-\epsilon^2/2}{1+\epsilon^2/2}$ we get a Lorentz
equivalent coordinates $\tilde{x}^\mu$ \cite{Sei,Bil1}
\begin{equation}
\left(\begin{array}{c}
	\tilde{x}^1\\ \tilde{x}^0
	\end{array}\right) = \frac{1}{\sqrt{1-\beta^2}}
		\left(\begin{array}{cc}
			1  &  -\beta \\
			-\beta  &  1
			\end{array}\right)
	\left(\begin{array}{c}
		x^1\\ x^0
		\end{array}\right) .
\end{equation}

In the new coordinate system $\tilde{x}^\mu$, it is convenient to
define the light-cone coordinates $x^{\pm} = (\tilde{x}^0 \pm
\tilde{x}^1)/\sqrt{2}$.
If $\epsilon$ is very small, the boost is very large and then the
periodicity of eq.(\ref{2.1}) becomes
\begin{equation}
  \left(\begin{array}{c}
	x^-\\ x^+
	\end{array} \right)
    \simeq  \left(\begin{array}{c}
			x^-\\ x^+
		\end{array} \right)
   + \left(\begin{array}{c}
		2\pi R\\ -\epsilon^2 \pi R
		\end{array}\right), \hspace{3ex}
  \tilde{x}^i \simeq \tilde{x}^i,\hspace{3ex}p^+=\frac{N}{R} +
	O(\epsilon^2). \label{2.3}
\end{equation}
(Here we should notice that the integer $N$ in (\ref{2.1}) agrees with
that in (\ref{2.3}).)
Then eq.(\ref{2.3}) clearly shows that in the $\epsilon
\rightarrow 0$ limit with the fixed $R$, an almost light-like circle
becomes exactly a light-like one with radius $R$, while the Lorentz
equivalent space-like circle (\ref{2.1}) has shrunk to zero size.
We get a discrete light-like coordinate system in this limit,
which we call the light-like limit (L$^3$).
Then we will be able to define DLCQ by quantizing theory on an almost
light-like circle or a vanishingly small space-like circle and taking
the L$^3$.
We should study whether this limit is really well-defined or not at
the quantum level.
Hence, following ref.\cite{Bil1,Bil2,Bil3}, we shall consider
amplitudes in field and string theories on $M_{25}\times S^1$ (radius
$R_s=\epsilon R$) and study their $\epsilon\rightarrow0$ limit.

%%%%%%%%%%%%%%%%%%%%%%%%%%%%%%%%%%%%%%%%%%%%%%%%%%%%%%%%%%%%%%%%%%%%%%
\section{The L$^3$ in field theory}
%%%%%%%%%%%%%%%%%%%%%%%%%%%%%%%%%%%%%%%%%%%%%%%%%%%%%%%%%%%%%%%%%%%%%%
In this section, we investigate the L$^3$ in field theory and review
the problem.
In order to make a clear comparison between field theory and string
theory, we take 26-dimensional $\phi^3$ scalar (tachyon) field
theory, where the mass square of a field is $-\frac{1}{\alpha'}$.

We consider four-point one-loop scattering processes, where each
external state has an incoming momentum $P_i\, (i=1,\cdots,4)$.
(see Figure 1.)
By using a parameter representation of the propagators,
$\Delta^{-1}=\int_0^1 dx\,x^{\Delta - 1}$, the scattering amplitude of
Figure 1(a) is given by\footnote{The signature of our metric is
$\eta_{\mu \nu}=diag (-1,1,1, \cdots,1)$ .},
%%%
\begin{figure}[htbp]
\centerline{\epsfbox{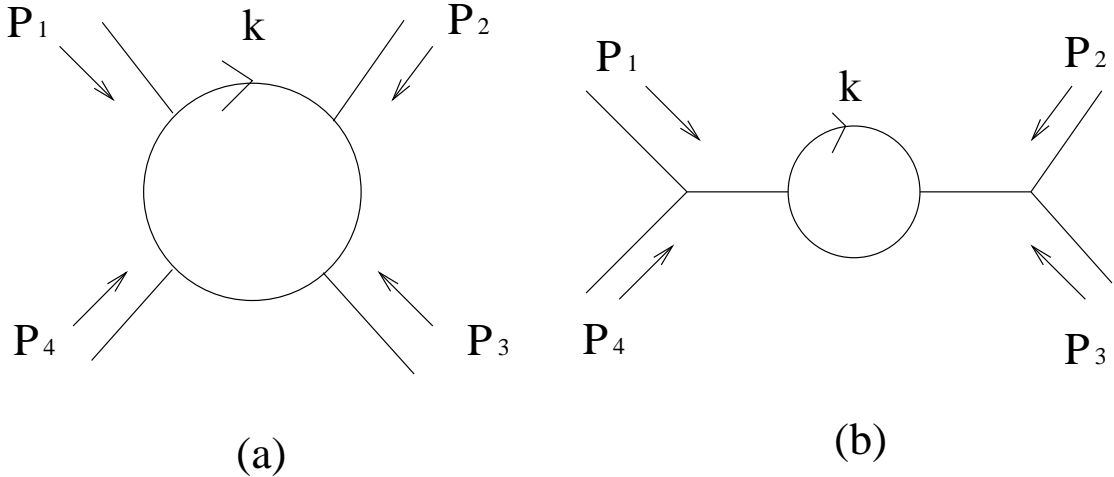}}
\caption{Two examples of four-point one-loop scattering processes in
field theory.}
\end{figure}
%%%
\begin{eqnarray}
A_{(a)}&=& \frac{g^4}{\alpha'^4} \int d^{26}k\,
  \frac{1}{k^2-\frac{1}{\alpha'}}\,
  \frac{1}{(k+P_2)^2-\frac{1}{\alpha'}}\,
  \frac{1}{(k+P_2+P_3)^2-\frac{1}{\alpha'}}\,
  \frac{1}{(k-P_1)^2-\frac{1}{\alpha'}}\nonumber\\
&=&g^4\int d^{26}k\, \int_{0}^{1} \prod_{i=1}^{4}dx_i\,
  x_{1}^{\alpha' k^2 -2}\, x_{2}^{\alpha' (k+P_2)^2 -2}\,
  x_{3}^{\alpha'(k+P_2+P_3)^2 -2}\,
  x_{4}^{\alpha'(k-P_1)^2-2}\label{3.1}.
\end{eqnarray}
Here we introduce the external dual momenta $p_i\,(i=1,\cdots,4)$ as
follows:
\begin{eqnarray}
	P_1&=&p_1-p_4,\label{3.2}\\
	P_2&=&p_2-p_1,\label{3.3}\\
	P_3&=&p_3-p_2,\label{3.4}\\
	P_4&=&p_4-p_3\label{3.5}.
\end{eqnarray}
Due to these definitions, the external momenta $P_i$ automatically
satisfy the conservation law ($\sum\limits_{i=1}^{4} P_i=0$).
Note that one of these $p_i$ is, in fact, a redundant degree of
freedom.
By using these dual momenta and shifting the loop momentum
$k\rightarrow -(k-p_1)$, we can rewrite (\ref{3.1}) as follows:
\begin{equation}
  A_{(a)}=g^4\int d^{26}k \int_{0}^{1} \prod_{i=1}^{4}dx_i
  \prod_{i=1}^{4} x_{i}^{-\alpha (-(k-p_i)^2) -1},\label{3.6}
\end{equation}
where $\alpha(s)$ is the Regge trajectory function of open string,
\begin{equation}
	\alpha (s) = \alpha' s +1. \label{3.7}
\end{equation}
Similarly, the scattering amplitude corresponding to Figure 1(b),
\begin{equation}
A_{(b)}=\frac{g^4}{\alpha'^4} \int d^{26}k~
  \frac{1}{(P_1 +P_4)^2 - \frac{1}{\alpha'}}\,
  \frac{1}{k^2 - \frac{1}{\alpha'}}\,
  \frac{1}{(P_1 +P_4 -k)^2 - \frac{1}{\alpha'}}\,
  \frac{1}{(P_1 +P_4)^2 - \frac{1}{\alpha'}},
\end{equation}
can be rewritten by
\begin{equation}
  A_{(b)}=g^4\int d^{26}k \int_{0}^{1} dx_1 dx_3 dz_2 dz_4~
	x_{1}^{-\alpha (-(k-p_1)^2) -1}\,
	x_{3}^{-\alpha (-(k-p_3)^2) -1}\,
	(z_{2}z_{4})^{-\alpha (-(p_1-p_3)^2) -1} \label{3.9}.
\end{equation}

Now we consider the L$^3$ of these scattering amplitudes, i.e.,
we give the amplitudes on $M_{25}\times S^1$ (radius $R_s=\epsilon R$)
and study their $\epsilon\rightarrow0$ limit.
On the $M_{25}\times S^1$ both the external and the loop
momenta along the compact direction are quantized as $p^1_i=n_i/R_s$,
$k^1=n/R_s$ and the integration over $k^1$ becomes a sum over $n$.
Then the amplitude $A_{(a)}$ in (\ref{3.6}) is given by,
\begin{eqnarray}
  A_{(a)}&=&g^4 \int d^{25} k'~\frac{1}{R_s} \sum_n
	\int_{0}^{1} \prod_{i=1}^{4}dx_i\,
	 (\prod_{i=1}^{4}x_{i})^{-2} \nonumber\\
  &&\times \exp \left[ \alpha'  \sum _{i=1}^{4}
	\left\{ \left( \frac{n - n_i}{R_s} \right)^2
        + (k'-p'_i)^2 \right\}
	\ln x_i \right],
\end{eqnarray}
where $(p'_i) = (p^0_i,p^2_i,p^3_i,\cdots,p^{25}_i)$ and
$(k')=(k^0,k^2,k^3,\cdots,k^{25})$.
Completing the squares for $n$ and $k'$, we can integrate over $k'$
and we get,
\begin{eqnarray}
  A_{(a)}&=& g^4
    \int_{0}^{1} \prod_{i=1}^{4}dx_i (\prod_{i=1}^{4}x_{i})^{-2}
    \left(\frac{\pi}{-\alpha' \ln (\prod\limits_{i=1}^{4} x_i)}
	\right)^{\frac{25}{2}}\nonumber \\
  &&\times \frac{1}{R_s} \sum_n ~\exp \left[ \alpha'
    \ln (\prod_{i=1}^{4} x_i) \left(
    \frac{n}{R_s} -\frac{\sum\limits_{i=1}^{4}n_i \ln x_i}
    {R_s \ln  (\prod\limits_{i=1}^{4} x_i)} \right)^2
    \right.\nonumber\\
  &&\hspace{3ex}+\left. \frac{\alpha'}{\ln
	(\prod\limits_{i=1}^{4} x_i)} \{\ln x_1 \ln x_2\,(p_1-p_2)^2
	+ \cdots + \ln x_3 \ln x_4\,(p_3-p_4)^2\}\right].\label{3.11}
\end{eqnarray}
We find that the $R_s$ dependent parts combine to give a
$\delta$-function in the L$^3$ \cite{Bil1,Bil3},
\begin{eqnarray}
  &&\lim_{\epsilon \rightarrow 0}\,\frac{1}{\epsilon R}\,
    \exp  \left[ \alpha' \ln (\prod_{i=1}^{4} x_i)\,
    \left(\frac{n}{\epsilon R} -\frac{\sum\limits_{i=1}^{4}n_i\ln x_i}
    {\epsilon R\,\ln (\prod\limits_{i=1}^{4} x_i)} \right)^2
    \right]\nonumber\\
  &&\quad =\left(\frac{\pi}{- \alpha' \ln
    (\prod\limits_{i=1}^{4}x_i)}\right)^{\frac{1}{2}}\,
    \delta\left(n-\frac{\sum\limits_{i=1}^{4}n_i \ln x_i}{\ln
    (\prod\limits_{i=1}^{4} x_i)}\right)\label{3.12}.
\end{eqnarray}
Note that $\ln (\prod\limits_{i=1}^{4} x_i) \leq 0$ since
$0\leq x_i\leq 1\,(i=1,\cdots,4)$.
Plugging the $\delta$-function into (\ref{3.11}) we can integrate over
one of the parameters $x_i$ and we can get a well-defined result which
agrees with that in the direct DLCQ \cite{Bil1}.
But this story breaks down if the $\delta$-function (\ref{3.12}) is
$\delta(0)$ identically and we encounter a pathology in the L$^3$
\cite{HP}.
Due to eq.(\ref{3.12}), it is obvious that this pathological
$\delta(0)$ appears iff $n_1=n_2=n_3=n_4=n$.
But this condition corresponds to the kinematical situation that
all the external momenta in the $S^1$ direction
$P_i^1=\frac{N_i}{R_s}\,(i=1,\cdots,4)$ are zero due to
eqs.(\ref{3.2})$\sim$(\ref{3.5}).
It is obvious that this pathology does not arise in this scattering
process of Figure 1(a).\footnote{See the footnote \ref{foot1}.}

Next we consider the L$^3$ of the amplitude $A_{(b)}$ (\ref{3.9}).
Similarly to eq.(\ref{3.11}), $A_{(b)}$  (\ref{3.9}) on the
$M_{25}\times S^1$ is given by,
\begin{eqnarray}
  A_{(b)}&=& g^4
	\int_{0}^{1} dx_1 dx_3 dz_2 dz_4
	 ( x_1 x_3)^{-2} (z_2 z_4)^{-\alpha(-(p_1-p_3)^2)-1 }
	\left( \frac{\pi}{-\alpha' \ln (x_1 x_3)}
	\right)^{\frac{25}{2}}\nonumber \\
  && \times \frac{1}{R_s} \sum_n ~
	\exp \left[ \alpha' \ln (x_1 x_3) \left(
	\frac{n}{R_s} -\frac{n_1 \ln x_1 + n_3 \ln x_3}
	{R_s \ln  (x_1 x_3)} \right)^2 \right.\nonumber\\
  && \left. \hspace{10ex} + \frac{\alpha'}{\ln (x_1 x_3) }
	\ln x_1 \ln x_3 (p_1-p_3)^2 \right] .
\end{eqnarray}
Then we get the following $\delta$-function in the L$^3$,
\begin{eqnarray}
  &&\lim_{\epsilon \rightarrow 0} \frac{1}{\epsilon R}
	\exp  \left[ \alpha' \ln (x_1 x_3) \left(
	\frac{n}{\epsilon R} -\frac{n_1 \ln x_1 +n_3 \ln x_3}
	{\epsilon R \ln(x_1x_3)} \right)^2  \right]\nonumber\\
  &&\quad =\left(\frac{\pi}{- \alpha'
	\ln ( x_1x_3)}\right)^{\frac{1}{2}}
	\delta\left(n-\frac{n_1\ln x_1+n_3\ln x_3}{\ln(x_1x_3)}\right) .
\end{eqnarray}
Here we see that $n_1 =n_3$ will give $\delta(0)$ in the amplitude.
This corresponds to the kinematical situation that
the external momenta $P_i^1 (=\frac{N_i}{R_s})$ satisfy
$\frac{N_1}{R_s} +\frac{N_4}{R_s}=0$ and  $\frac{N_2}{R_s}
+\frac{N_3}{R_s}=0 $ due to eqs.(\ref{3.2})$\sim$(\ref{3.5})
and this means that any external momenta $P_i^1$ do not flow through
the loop.
In this case when $n=n_1$ in the sum over the discrete loop momentum,
the divergence appears. In the original loop momentum language,
this divergence is obviously due to the zero-mode loop since we have
shifted the loop momentum $k\rightarrow -(k-p_1)$.
Hence this pathology is the same as before \cite{HP}.

In field theory, we must add all possible Feynman diagrams as Figure
1(a) and 1(b) to calculate scattering amplitudes. By the above
consideration, it is understood that in the L$^3$ only the zero-mode
loop contributions among them are divergent as $\delta(0)$.
It is well known that in the direct DLCQ the zero-modes are not
dynamical degrees of freedom and they are written by the non-zero
modes due to a nonlinear operator equation (zero-mode constraint
\cite{MY}). Thus the zero-mode loop contributions do not appear in the
direct DLCQ.

%%%%%%%%%%%%%%%%%%%%%%%%%%%%%%%%%%%%%%%%%%%%%%%%%%%%%%%%%%%%%%%%%%%%%%%
\section{Multi-loop scattering amplitudes and their L$^3$ in string
	theory}\label{secMLST}
%%%%%%%%%%%%%%%%%%%%%%%%%%%%%%%%%%%%%%%%%%%%%%%%%%%%%%%%%%%%%%%%%%%%%%%
In this section we investigate the L$^3$ of the multi-loop scattering
amplitudes in bosonic string theory.
In order to investigate the zero-mode loop problem,
we adopt the method of Kikkawa, Sakita and Virasoro \cite{KSV} to
construct scattering amplitudes since the integration over loop
momenta are explicitly kept in the amplitudes.
In this method, s-t channel dualities and crossing symmetries are
explicitly taken into account and it is easy to investigate whether we
can take the L$^3$ of these amplitudes or not. First we briefly review
their method. We clarify the rules for self-crossing lines in loop
amplitudes and give tachyon scattering amplitudes in bosonic string
theory. Then we will show that the L$^3$ of these amplitudes are
well-defined.
%%%%%%%%%%%%%%%%%%%%%%%%%%%%%%%%%%%%%%%%%%%%%%%%%%%%%%%%%%%%%%%%%%%%%%%
\subsection{KSV method in bosonic open string theory}
%%%%%%%%%%%%%%%%%%%%%%%%%%%%%%%%%%%%%%%%%%%%%%%%%%%%%%%%%%%%%%%%%%%%%%%
In this subsection, we review the method of Kikkawa, Sakita and Virasoro
in bosonic open string theory \cite{KSV}.\footnote{For simplicity, we
restrict ourselves to the four-point planer amplitude in this
subsection. We will extend this result $N$-point ones in Appendices A
and B.} And then we will extend this method to bosonic closed string
theory in the next subsection.

\begin{figure}[htbp]
\centerline{\epsfbox{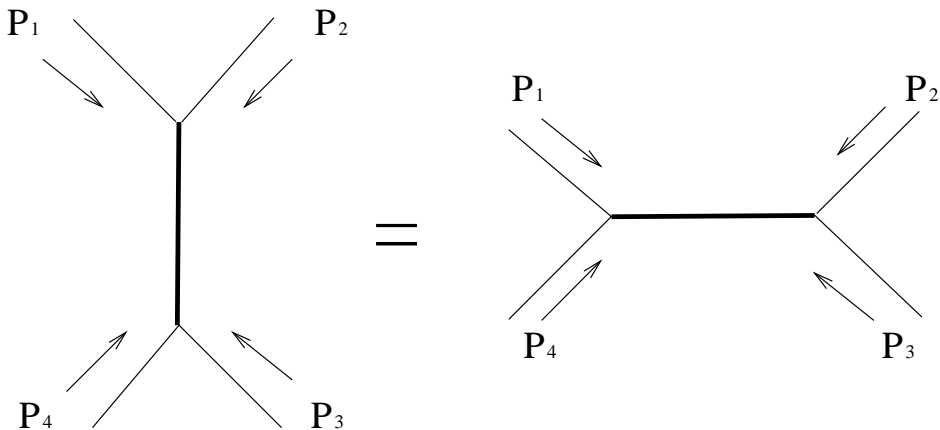}}
\caption{s-t channel duality in the Veneziano amplitude.}\label{figVA}
\end{figure}

\begin{figure}[htbp]
\centerline{\epsfbox{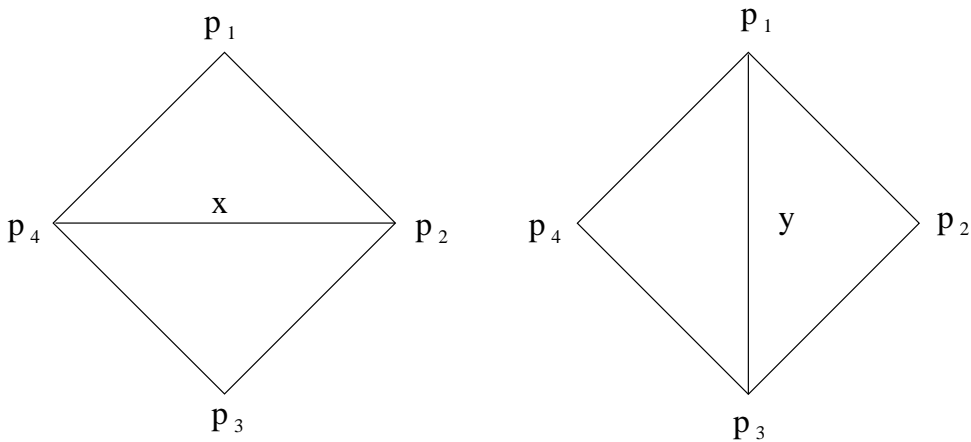}}
\caption{Two dual diagrams for the Veneziano amplitude.}\label{figDG}
\end{figure}

As is well known, the tachyon four-point tree amplitude
(Veneziano amplitude) is given by,
\begin{equation}
  A_4^{(0)} = g^2 \int _{0}^{1} dx\,
	x^{-\alpha(s)-1}y(x)^{-\alpha(t)-1}  ,\label{4.1}
\end{equation}
where
\begin{eqnarray}
	y(x)&=&1-x\label{4.2},\\
	s&=&-(P_1 +P_2)^2,\label{4.3}\\
        t&=& -(P_2+P_3)^2,\label{4.4}
\end{eqnarray}
and $\alpha(s)$ is given in eq.(\ref{3.7}) and $P_i$
($P_i^2=\frac{1}{\alpha'}, \,\, i=1,\cdots,4$) are
incoming external tachyon momenta. This amplitude manifestly has
s-t channel duality (Figure \ref{figVA}). Here we consider the dual
diagram to each diagram in Figure \ref{figVA}; i.e., we consider the
quadrilateral defined by four sides which are dual to four external
lines and two diagonals which are dual to the s- and t-channel
propagators, respectively (Figure \ref{figDG}).
Since the parameter of a propagator, e.g., $y$, corresponds to a line
in the dual diagram, we simply call it the line $y$. We assign dual
external momenta $p_i$ ($i=1,\cdots,4$), which are defined by
eqs.(\ref{3.2})$\sim$(\ref{3.5}), to the four vertices of the
quadrilateral in Figure \ref{figDG}. With these dual momenta,
eqs.(\ref{4.3}) and (\ref{4.4}) are rewritten by,
\begin{eqnarray}
	s&=& -(p_2-p_4)^2,\\
        t&=&-(p_1-p_3)^2.
\end{eqnarray}

\begin{figure}[htbp]
\epsfxsize=110mm
\centerline{\epsfbox{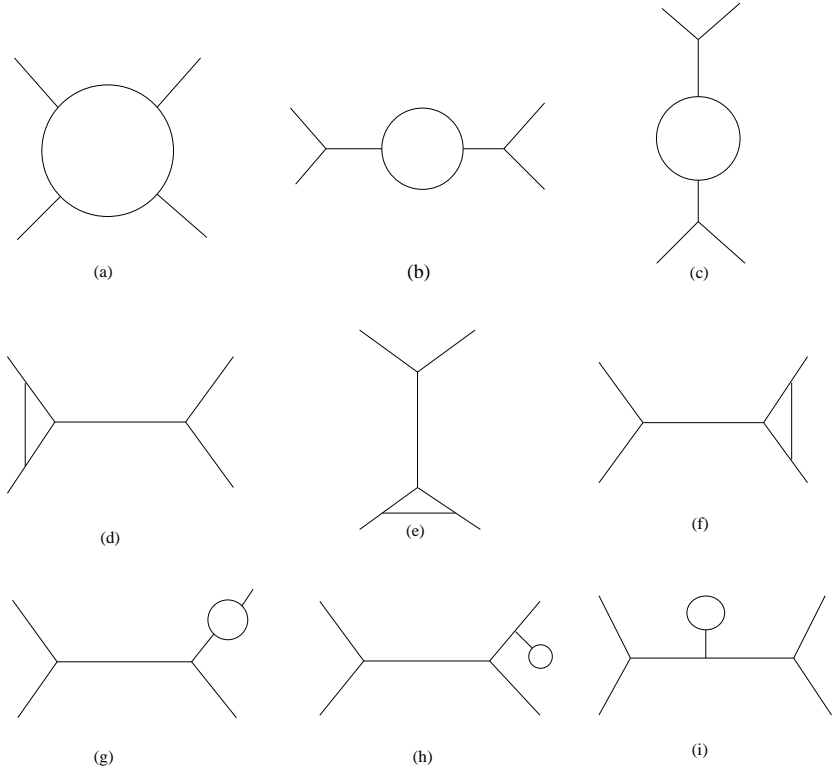}}
\caption{Examples of various one-loop field theoretical Feynman diagrams
that are connected by s-t channel duality.}
\label{figFD}
\end{figure}

Next we consider the four-point one-loop amplitude. The amplitude
should contain the resonances in various field theoretical Feynman
diagrams which are connected by s-t channel duality (Figure
\ref{figFD}).
%%%
%%%
\begin{figure}[htbp]
\epsfxsize=110mm
\centerline{\epsfbox{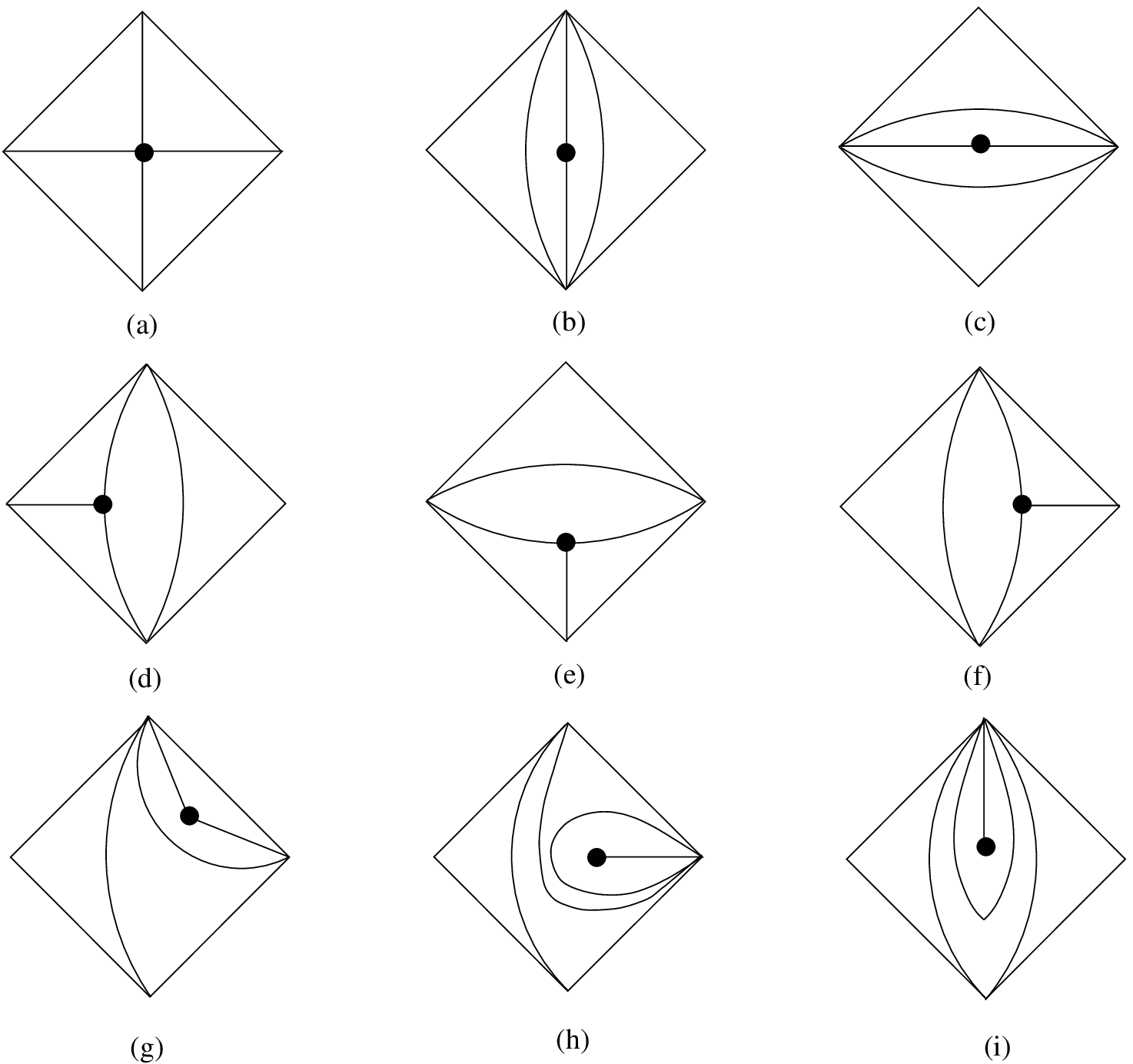}}
\caption{Examples of various dual diagrams for the one-loop field
 theoretical Feynman diagrams in Figure \ref{figFD}.}
\label{figDGOL}
\end{figure}

\begin{figure}[htbp]
\centerline{\epsfbox{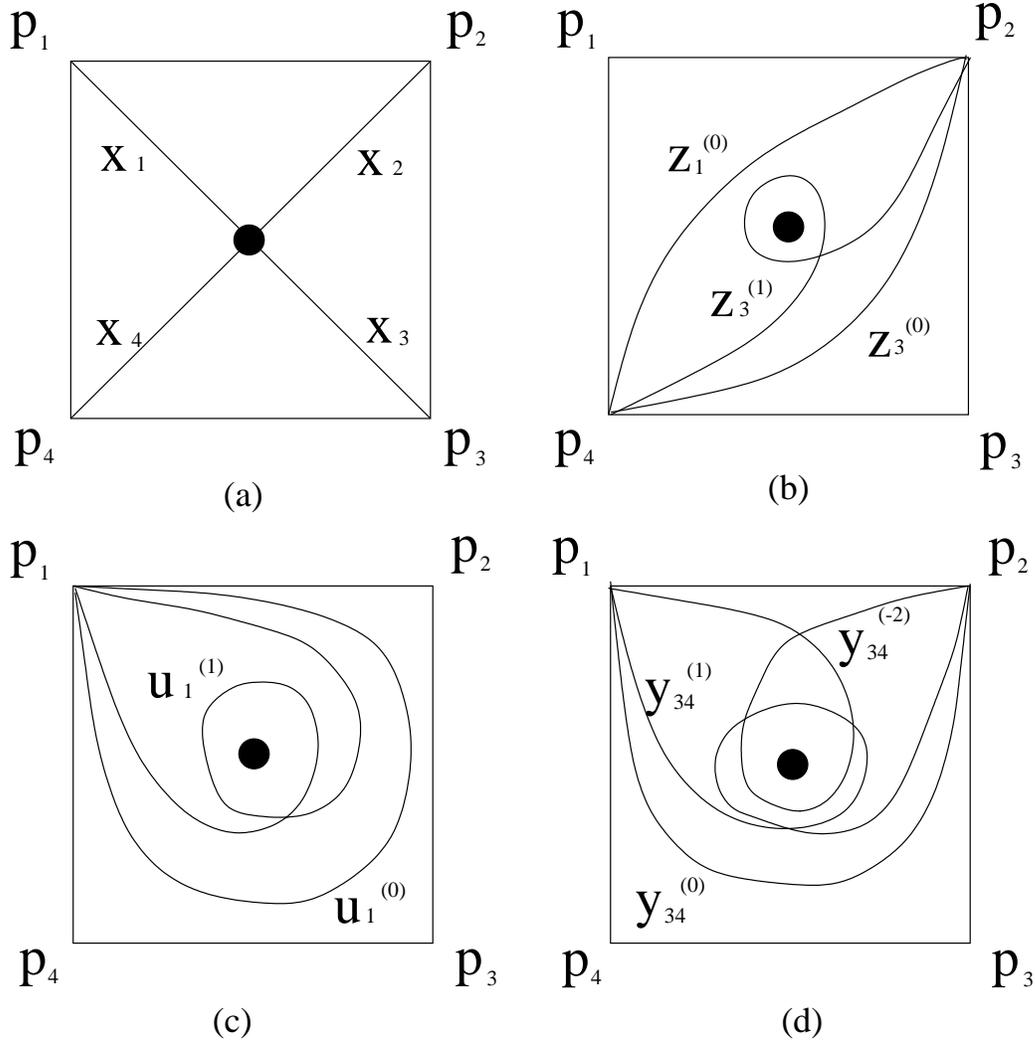}}
\caption{Some lines which must be added in one-loop dual diagrams;
$y_{34}^{(0)}$ corresponds to a self-energy correction to the external
line $P_2$, and $u_1^{(0)}$ is a tadpole line. Here, we should notice
that the line $y_{34}^{(-1)}$ corresponds to the external line
 $P_2$ itself. Therefore, it is not included in the amplitude
(\ref{4.8}). } \label{figOLSC}
\end{figure}

In order to construct the one-loop amplitude, we also consider
the dual diagrams to those in Figure \ref{figFD}; i.e., we consider the
quadrilateral defined by four sides which are dual to four external
lines and an internal point which is dual to a closed loop and all
possible lines which can be drawn between any two points of this
quadrilateral, which are dual to all possible propagators (Figure
\ref{figDGOL}).
In each dual diagram, we always have four lines which connect two
points of four vertices and one internal point and the four lines
correspond to four propagators.  Two lines which cross each other
correspond to channels where we can not simultaneously find
resonances. They have the same relation as the s-t channel in the
Veneziano amplitude. Here, we can write the one-loop amplitude
schematically as (\ref{4.1}):
\begin{equation}
  A_4^{(1)} \sim g^4 \int d^{26}k \int_0^1 \prod_{i=1}^{4}d x_i
  \prod_{j=1}^{k}y_j(x_1,\cdot \cdot \cdot,x_4)^{-\alpha(s_j)-1}
   ,\label{4.5}
\end{equation}
where factors $y_j(x_1,\cdot \cdot \cdot,x_4)$ are
parameters of all possible propagators corresponding to all possible
lines, which join pairs of points and do not cross themselves in dual
diagrams and four of those are independent parameters $x_i\,
(i=1,\cdots,4)$ and $-s_j$ are square of momenta carried by the j-th
propagator.
But this is not the end of the story.
In order to obtain the correct one-loop amplitude,
we must add the contributions of lines which cross themselves in dual
diagrams, e.g., $z_3^{(1)}$ in Figure \ref{figOLSC} (b), which do not
correspond to any propagators in the field theoretical
Feynman diagrams \cite{KSV,BHS}.
Hence, the $k$ factors $y_j(x_1,\cdot \cdot \cdot,x_4)$ should
correspond to all possible topologically inequivalent lines
that join pairs of points in dual diagrams.
Since the $k$ lines are not independent and we choose four
independent lines $x_i\, (i=1,\cdots,4)$ in Figure \ref{figOLSC}
(a). Then all other lines
$z_i^{(n)},\, u_i^{(n)} \,(i=1,\cdots,4 ,\,n\geq0)$
and $y_{12}^{(n)},\,y_{23}^{(n)},\,y_{34}^{(n)},\,y_{41}^{(n)}\,
(n \geq 0,\,n \leq -2)$ are functions of $x_i$ as $y(x)$ in
eq.(\ref{4.2}). We draw some dependent lines in Figure
\ref{figOLSC} (b),(c),(d).
We can determine these functions and  we obtain
the one-loop amplitude as follows:
\begin{eqnarray}
  A_4^{(1)}&=& g^4 \int d^{26}k \int_{0}^{1} \prod_{i=1}^{4} dx_i\,
	\rho(x_i) \prod_{i=1}^{4}\, x_i^{-\alpha(-(k-p_i)^2)-1}
    \left(\prod_{n=0}^{\infty} z_1^{(n)}z_3^{(n)}
        \right)^{-\alpha(-(p_2-p_4)^2)-1} \nonumber \\
  && \times
    \left(\prod_{n=0}^{\infty}z_2^{(n)}z_4^{(n)}
	\right)^{-\alpha(-(p_1-p_3)^2)-1 }\,
    \left(\prod_{n=0}^{\infty} u_1^{(n)} u_2^{(n)} u_3^{(n)}
	u_4^{(n)}\right)^{-\alpha (0) -1} \nonumber\\
  && \times
	\left(\prod_{n=0}^{\infty} y_{12}^{(n)} y_{23}^{(n)}
	y_{34}^{(n)} y_{41}^{(n)} \, \prod_{n=2}^{\infty}
	y_{12}^{(-n)} y_{23}^{(-n)} y_{34}^{(-n)}
	y_{41}^{(-n)}\right)^{-\alpha (-\frac{1}{\alpha'}) -1},  \label{4.8}
\end{eqnarray}
where $\rho(x_i)$ is the measure of the integral, which will be
determined later. Note that the power of each line depends on the
momentum of the corresponding propagator but it does not depend on
$n$, i.e., how many times the line crosses itself.
We should notice that if we put all the dependent parameters
$z_i^{(n)}(x_i),\,u_i^{(n)}(x_i),\,y_{ij}^{(n)}(x_i)$ and measure
$\rho(x_i)$ equal to one, this amplitude (\ref{4.8}) is reduced to the
one in the field theory (\ref{3.6}).

\begin{figure}[htbp]
\centerline{\epsfbox{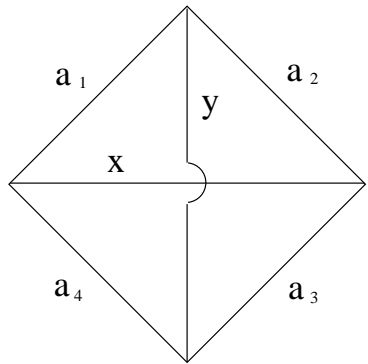}}
\caption{A part of the duality diagram. Line $y$ is determined by
 $a_1,\,a_2,\,a_3,\,a_4$ and $x$.}
\label{figXY}
\end{figure}

Next we determine the forms of the functions,
$z_i^{(n)}(x_i), u_i^{(n)}(x_i), y_{ij}^{(n)}(x_i)$ as $y(x)$ in
eq.(\ref{4.2}).
For example, the lines $x$ and $y$ in eq.(\ref{4.2}) are two diagonals
that cross each other in the quadrilateral whose four sides correspond
to external lines.
Here we assume that even if four sides of the quadrilateral do not
correspond to external lines, i.e., four sides have the parameters
$a_i \,(i=1,\cdots,4)$ (Figure \ref{figXY}),
one diagonal $y$ is described by the
function of the other diagonal $x$ and four
sides $a_i \,(i=1,\cdots,4)$ as follows:
\begin{equation}
	y=f(x; a_1,a_2,a_3,a_4)\label{4.9}.
\end{equation}
Of course, if all the sides of the quadrilateral correspond to
the external lines; i.e.,  $a_1=a_2=a_3=a_4=0$, eq.(\ref{4.9})
should be reduced to (\ref{4.2}).
Under this assumption, for example, the line $z_1^{(0)}$
in Figure \ref{figOLSC} (b), which is a diagonal of the quadrilateral
whose sides are $(x_4,\,0,\,0,\,x_2)$, is described by\footnote{A line
may be regarded as the diagonal of different quadrilaterals.
Uniqueness of the expression of a diagonal in this prescription is
proved in the appendix B of \cite{KSV}.},
\begin{equation}
	z_1^{(0)}=f(x_1 ; x_{4},0,0,x_{2}).
\end{equation}
And $u_1^{(0)}$ and $y_{34}^{(0)}$, in Figure \ref{figOLSC}
(c), (d) are also given by,
\begin{eqnarray}
	u_1^{(0)}&=&f(x_3 ; x_{1},z_2,z_4,x_{1}),\\
	y_{34}^{(0)}&=&f(x_4 ; x_{2},z_3,0,x_{1}).
\end{eqnarray}
Furthermore, we assume that $n$ times winding (self-crossing) lines,
$z_i^{(n)}$, $u_i^{(n)}$ and $y_{ij}^{(n)}$, are essentially the
same diagonals as $z_i^{(0)}$, $u_i^{(0)}$ and $y_{ij}^{(0)}$,
respectively, but since they cross all the $x_i (i=1,\cdots,4)$ $n$
times, the corresponding diagonals, $x_1, x_3$ and $x_4$,
should be multiplied by $(x_1x_2x_3x_4)^n$. Then we have
\begin{eqnarray}
	z_1^{(n)}&=&f(x_1w^n ; x_{4},0,0,x_{2}),\label{4.13}\\
	u_1^{(n)}&=&f(x_3w^n ; x_{1},z_2,z_4,x_{1}),\\
	y_{34}^{(n)}&=&f(x_4w^n ; x_{2},z_3,0,x_{1}),\\
	w&\equiv&x_1x_2x_3x_4.
\end{eqnarray}

Now the remaining problem is to determine one function
$y=f(x;a_1,a_2,a_3,a_4 )$ \cite{KSV}. In fact,
we can determine this function by comparing the result of
$N \,(\geq8)$-point tree amplitude in the operator formalism with that
by the KSV method. In appendix A, we show this in some details.
The function $y=f(x; a_1,a_2,a_3,a_4 )$ is given by \cite{KSV},
\begin{equation}
  y=f(x;a_1,a_2,a_3,a_4)=
    \frac{1-x\alpha_2\alpha_3}{1-x\alpha_2 \alpha_3 a_1}\,
    \frac{1-x\alpha_2\alpha_3a_1a_4}{1-x\alpha_2\alpha_3a_4},\label{4.17}
\end{equation}
where
\begin{eqnarray}
  a_2&=&\frac{1-\alpha_2}{1-\alpha_2 a_1}\,
	\frac{1-\alpha_2 a_1 x}{1-\alpha_2 x}\,,\\
  a_3&=&\frac{1-\alpha_3}{1-\alpha_3 a_4}\,
	\frac{1-\alpha_3 a_4 x}{1-\alpha_3 x}\,.\label{4.19}
\end{eqnarray}
Here we should notice that when all sides are external lines,  $a_i =
0 $, eq.(\ref{4.17}) is reduced to $y = f(x;0,0,0,0) =1-x$.

Since the function has been determined, the one-loop scattering
amplitude (\ref{4.8}) is also determined up to the measure function
$\rho(x_i)$. By the requirement of crossing symmetry in the amplitude
(\ref{4.8}), we can determine $\rho(x_i)$ up to an invariant function
\cite{KSV}:
\begin{equation}
  \rho(x_i)\propto\frac{1}{(1-x_1x_2)(1-x_2x_3)(1-x_3x_4)(1-x_4x_1)}.
\end{equation}
In appendix B, by comparing the result of one-loop amplitudes in the
operator formalism with those by the KSV method, we have determined the
invariant function and then $\rho(x_i)$ is given by,
\begin{eqnarray}
  \rho(x_i)&
	=&\frac{1}{(1-x_1x_2)(1-x_2x_3)(1-x_3x_4)(1-x_4x_1)}
	\left[f(w)\right]^{-24}\,,\label{4.21}\\
  f(w)&\equiv&\prod_{n=1}^{\infty} (1- w^n)\,.
\end{eqnarray}

In conclusion, by this KSV method, we can construct the one-loop
scattering amplitude, which is explicitly s-t channel dual and
crossing symmetric by construction, up to $\left[f(w)\right]^{-24}$
factor. In appendix B, we show that the $N$-point one-loop amplitudes
constructed by this KSV method agree with those in the operator
formalism.

  From the results of four-point tree and one-loop amplitudes,
the extension to the four-point $N$-loop amplitude is obvious.
In the $N$-loop case, we should consider the dual quadrilateral which
has $N$ internal points and choose $3N + 1$ independent lines
$x_j\,(j=1,\cdots,3N+1)$ among all the possible topologically
inequivalent lines which connect any two points in the dual diagrams.
Then the general form of four-point $N$-loop amplitude is given by,
\begin{eqnarray}
  A_4^{(N)} &=& g^{2N+2} \int \prod_{J=1}^{N} d^{26}k_J
	\int_0^1 \prod_{j=1}^{3N+1}dx_j\, G \,
	D^{-\alpha(-(p_1-p_3)^2)-1}\,
	C^{-\alpha(-(p_2-p_4)^2)-1}\nonumber\\
  &&\times
	(\prod_{J=1}^{N} A_{J}^{-\alpha(-(k_J-p_2)^2)-1})\,
	(\prod_{J=1}^{N} A_{J}'^{-\alpha(-(k_J-p_4)^2)-1})\,
	(\prod_{J=1}^{N} B_{J}^{-\alpha(-(k_J-p_1)^2)-1})\nonumber\\
  &&\times
	(\prod_{J=1}^{N} B_{J}'^{-\alpha(-(k_J-p_3)^2)-1})\,
	(\prod_{J < K} X_{JK}^{-\alpha(-(k_J-k_K)^2)-1})\,,\label{4.23}
\end{eqnarray}
where $A_J$, $A'_J$, $B_J$, $B'_J$ and $X_{JK}\,(J,K =1,\cdots,N)$
are the products of all parameters, dependent as well as independent,
which correspond to topologically inequivalent lines connecting the
internal point of the loop momentum $k_J$ with those of momenta,
$p_2,p_4,p_1,p_3$ and $k_K$, respectively.
Furthermore $C$ is the product of lines between $p_2$ and $p_4$ and
$D$ is the product of those between $p_1$ and $p_3$.
$G$ is a function of the parameters $x_j$ and it is independent of
both internal $k_J \,(J=1,\cdots,N)$ and external
$p_i\,(i=1,\cdots,4)$ momenta.

%%%%%%%%%%%%%%%%%%%%%%%%%%%%%%%%%%%%%%%%%%%%%%%%%%%%%%%%%%%%%%%%%
\subsection{KSV method in bosonic closed string theory}
%%%%%%%%%%%%%%%%%%%%%%%%%%%%%%%%%%%%%%%%%%%%%%%%%%%%%%%%%%%%%%%%%
The KSV method is applicable to the bosonic closed string.
In closed string amplitudes, we should change all the real parameters
of propagators in open string amplitudes for complex ones and sum up
all the terms corresponding to the different orderings of the external
lines. For example, the closed string four-point tree amplitude
(Shapiro-Virasoro amplitude), which corresponds to (\ref{4.1}) in open
string, is given by,
\begin{eqnarray}
  I_4^{(0)} &=& \frac{\kappa^2}{4\pi} \int_{|x| \leq 1} d^2x\,
	|x|^{-\alpha_C(s)-2}\,|1-x|^{-\alpha_C(t)-2}
	+ (P_2\leftrightarrow P_3)\label{4.24}\\
  &=& \frac{\kappa^2}{4\pi} \int_{\bf C} d^2x\,
	|x|^{-\alpha_C(s)-2}\,|1-x|^{-\alpha_C(t)-2},
\end{eqnarray}
where $P_i\,\,(P_i^2=\frac{4}{\alpha'}\,,\,i=1,\cdots,4)$ are incoming
external tachyon momenta and the Regge trajectory function $\alpha_C$
of closed string is given by,
\begin{equation}
	\alpha_C(s)=\frac{\alpha'}{2}s-2.
\end{equation}
In eq.(\ref{4.24}) the first and second terms
correspond to the different orderings of the external lines and
by combining these two terms we can obtain the complete amplitude
which has s-t-u channel duality \cite{GSW}.
For simplicity, however, we will henceforth consider only the term which
correspond to the ordering $P_1\rightarrow P_2\rightarrow
P_3\rightarrow P_4$ as the first term in eq.(\ref{4.24}).

By the above considerations, closed string four-point one-loop
amplitude, which corresponds to (\ref{4.8}), is given by,
\begin{eqnarray}
I_4^{(1)}&=&\left(\frac{\kappa}{4 \pi}\right)^4 \int d^{26}k\,
	\int_{|x_i| \leq 1} \prod_{i=1}^{4} d^2x_i\,
	|\rho(x_i)|\,
	\prod_{i=1}^{4} |x_i|^{-\alpha_C(-(k-p_i)^2)-2}
        \left|\prod_{n=0}^{\infty}z_1^{(n)}\,
	z_3^{(n)}\right|^{-\alpha_C(-(p_2-p_4)^2)-2}\nonumber\\
  &&\times\,
	\left|\prod_{n=0}^{\infty}\,
	z_2^{(n)}z_4^{(n)}\right|^{-\alpha_C(-(p_1-p_3)^2)-2}\,
	\left|\prod_{n=0}^{\infty} u_1^{(n)} u_2^{(n)} u_3^{(n)}
	u_4^{(n)}\right|^{-\alpha_C (0) -2}\nonumber\\
  &&\times\,
	\left|\prod_{n=0}^{\infty} y_{12}^{(n)} y_{23}^{(n)} y_{34}^{(n)}
	y_{41}^{(n)} \prod_{n=2}^{\infty}
	y_{12}^{(-n)} y_{23}^{(-n)} y_{34}^{(-n)}
	y_{41}^{(-n)}\right|^{-\alpha_C (-\frac{1}{\alpha'}) -2}\,,
	\label{eqCFO}
\end{eqnarray}
where
\begin{equation}
	|\rho(x_i)|
         =\frac{1}{|(1-x_1x_2)(1-x_2x_3)(1-x_3x_4)(1-x_4x_1)|^2}\,
         |f(w)|^{-48},
\end{equation}
which corresponds to (\ref{4.21}).
The complex parameters $z_i^{(n)},\,u_i^{(n)}$ and $y_{ij}^{(n)}$ are
given by the same functions of the complex parameters $x_i$ as in
(\ref{4.13})$\sim$(\ref{4.19}) .
In fact, it is easy to see that this one-loop amplitude agrees with
that in the operator formalism in closed string theory.
Furthermore, the closed string four-point $N$-loop amplitude, which
corresponds to (\ref{4.23}), is given by,
\begin{eqnarray}
  I_4^{(N)}&=&\frac{\kappa^{2N+2}}{(4\pi)^{3N+1}}
	 \int \prod_{J=1}^{N} d^{26}k_J
	\int_{|x_j| \leq 1} \prod_{j=1}^{3N+1}d^2x_j\,|G|
	|D|^{-\alpha_C(-(p_1-p_3)^2)-2}
	|C|^{-\alpha_C(-(p_2-p_4)^2)-2}\nonumber\\
  &&\times\,(\prod_{J=1}^{N} |A_{J}|^{-\alpha_C(-(k_J-p_2)^2)-2})\,
	(\prod_{J=1}^{N} |A_{J}^{'}|^{-\alpha_C(-(k_J-p_4)^2)-2})\,
	(\prod_{J=1}^{N} |B_{J}|^{-\alpha_C(-(k_J-p_1)^2)-2})\nonumber\\
  &&\times\,(\prod_{J=1}^{N} |B_{J}^{'}|^{-\alpha_C(-(k_J-p_3)^2)-2})\,
	(\prod_{J<K} |X_{JK}|^{-\alpha_C(-(k_J-k_K)^2)-2})\label{4.29}\,,
\end{eqnarray}
where the complex functions $A_J, A'_J, B_J, B'_J, C, D, G$ and
$X_{JK}\,(J,K =1,\cdots,N)$ exactly correspond to those in open
string, respectively.
%%%%%%%%%%%%%%%%%%%%%%%%%%%%%%%%%%%%%%%%%%%%%%%%%%%%%%%%%%%%%%%%%%%%
\subsection{The L$^3$ of multi-loop scattering amplitudes}
%%%%%%%%%%%%%%%%%%%%%%%%%%%%%%%%%%%%%%%%%%%%%%%%%%%%%%%%%%%%%%%%%%%%
In this subsection, we take the L$^3$ of amplitudes (\ref{4.23}) and
(\ref{4.29}), i.e., we give the amplitudes on $M_{25}\times S^1$
(radius of $S^1$ is $R_s=\epsilon R$) and study their $\epsilon
\rightarrow 0$ limit.

First, we consider the L$^3$ in open string theory.
On the $M_{25}\times S^1$ both the external and the loop
momenta along the compact direction are quantized as
$p_i^1=n_i/R_s , \, k_J^1=n^{(J)}/R_s \,\,\,
(i=1,\cdots ,4, \,\, J=1,\cdots ,N)$ and the integration over $k_J^1$
becomes the sum over $n^{(J)}$, respectively. Then the amplitude
$A_4^{(N)}$ in (\ref{4.23}) is given by,
\begin{eqnarray}
  A_4^{(N)} &=& g^{2N+2} \left(\frac{1}{R_s}\right)^N
	\sum_{n^{(1)}, \cdots ,n^{(N)}}
                 \int \prod_{J=1}^{N} d^{25}k'_J
	\int_0^1 \prod_{j=1}^{3N+1} dx_j\, G \,
	D^{-\alpha(-(p_1-p_3)^2)-1}\,
	C^{-\alpha(-(p_2-p_4)^2)-1}  \nonumber \\
  &&\times \prod_{J=1}^N \bigg(A_J A'_J B_J B'_J
	\prod_{K=J+1}^N X_{JK}\bigg)^{-2}
 \exp \left[ \alpha' \sum_{J=1}^{N}
        \Biggl\{ \left(\left( k_{J}^1 -p_2^1 \right)^2
       + (k'_J-p_2')^2 \right)\ln A_J    \right. \nonumber\\
    &&+ \left( \left( k_{J}^1 -p_4^1 \right)^2
       + (k'_J-p_4')^2 \right)\ln A'_J
       +\left( \left( k_{J}^1 -p_1^1\right)^2
       + (k'_J-p_1')^2 \right)\ln B_J \\
    &&+ \left.
         \left( \left( k_{J}^1 -p_3^1\right)^2
       + (k'_J-p_3')^2 \right)\ln B'_J
       + \sum_{K=J+1}^N \left( \left(k_{J}^1 -k_K^1 \right)^2
       + (k'_J-k'_K)^2 \right) \ln X_{JK}  \Biggr\} \right], \nonumber
\end{eqnarray}
where $(p'_i )=( p_i^0 , p_i^2 , p_i^3, \cdots ,p_i^{25})$ and
$(k'_J)=(k^0_J, k_J^2, k_J^3 ,\cdots ,k_J^{25}) $.
Just as we completed the squares for $n$ and $k'$ in section 3, we
successively complete the squares for $N$ loop momenta, $n^{(J)}$ and
$k'_J$, starting with $n^{(N)}$ and $k'_N$, then $n^{(N-1)}$ and
$k_{N-1}$, etc.. We can integrate over $k'_J$, however, in order to
investigate whether we can take the L$^3$ of this amplitude or not,
henceforth we only concentrate on the following $R_s$ dependent part
$S_o$ of the amplitude,
\begin{eqnarray}
S_o&=& \left(\frac{1}{R_s} \right)^N  \sum_{n^{(1)}, \cdots ,n^{(N)}}
   \exp \left[ \alpha' \sum_{J=1}^{N}
        \left\{\left(\frac{n^{(J)}}{R_s} - \frac{n_2}{R_s} \right)^2
        \ln A_J
       +\left(  \frac{n^{(J)}}{R_s} -\frac{n_4}{R_s} \right)^2
        \ln A'_J \right. \right. \nonumber\\
    &&\hspace{5ex}+\left(\frac{n^{(J)}}{R_s}-\frac{n_1}{R_s}\right)^2
        \ln B_J+\left( \frac{n^{(J)}}{R_s} -\frac{n_3}{R_s} \right)^2
        \ln B'_J \nonumber\\
    &&\hspace{10ex}\left.\left.+ \sum_{K=J+1}^{N}
	 \left(\frac{n^{(J)}}{R_s}-\frac{n^{(K)}}{R_s}\right)^2
	\ln X_{JK}  \right\}\right]. \label{4.31}
\end{eqnarray}
Completing the square for $n^{(N)}$, we obtain,
\begin{eqnarray}
S_o&=& \left( \frac{1}{R_s} \right)^N
      \sum_{n^{(1)},\cdots,n^{(N)}}
  \exp \left[ \alpha'\ln \left(A_N A'_N B_N B'_N
      \prod_{K=1}^{N-1} X_{KN}\right) \right.\nonumber\\
  &&\times\Bigg(\frac{n^{(N)}}{R_s} -
   \frac{n_2 \ln A_N + n_4 \ln A'_N +n_1 \ln B_N + n_3 \ln B'_N
    + \sum\limits_{K=1}^{N-1} n^{(K)} \ln X_{KN}}
    {R_s \ln \left(A_N A'_N B_N B'_N \prod\limits_{K=1}^{N-1}
    X_{KN}\right)} \Bigg)^2  \Bigg] \nonumber\\
  &&\times\exp \Biggl[  \alpha'  \sum_{J=1}^{N-1}
    \Biggl\{ \left( \frac{n^{(J)}}{R_s} - \frac{n_2}{R_s} \right)^2
    \ln A^{(N-1)}_J+\left(\frac{n^{(J)}}{R_s}-\frac{n_4}{R_s}\right)^2
    \ln A'^{(N-1)}_J \nonumber \\
  &&\hspace{10ex} +\left(  \frac{n^{(J)}}{R_s} -\frac{n_1}{R_s} \right)^2
    \ln B^{(N-1)}_J  +\left( \frac{n^{(J)}}{R_s} -\frac{n_3}{R_s}
    \right)^2 \ln B'^{(N-1)}_J \nonumber\\
  &&\hspace{10ex}+ \sum_{K=J+1}^{N-1} \left(\frac{n^{(J)}}{R_s}-
	\frac{n^{(K)}}{R_s} \right)^2  \ln X^{(N-1)}_{JK}\Biggr\}
	+ \cdots \Biggr], \label{4.32}
\end{eqnarray}
where $A_J^{(N-1)}, A'^{(N-1)}_J, B_J^{(N-1)}, B'^{(N-1)}_J
\,(1\leq J\leq N-1)$ and $X_{IJ}^{(N-1)}\,(1\leq I<J\leq N-1)$ are
given by,
\begin{eqnarray}
  \ln A^{(N-1)}_J  &=& \ln A_J  + \frac{\ln A_N \ln X_{JN}}
	{\ln \left(A_N A'_N B_N B'_N \prod\limits_{K=1}^{N-1}
	X_{KN}\right)}~, \label{4.33}\\
  \ln A'^{(N-1)}_J &=& \ln A'_J + \frac{\ln A'_N \ln X_{JN}}
	{\ln \left(A_N A'_N B_N B'_N \prod\limits_{K=1}^{N-1}
	X_{KN}\right)}~,\label{4.34}\\
  \ln B^{(N-1)}_J  &=& \ln B_J  + \frac{\ln B_N \ln X_{JN}}
	{\ln \left(A_N A'_N B_N B'_N \prod\limits_{K=1}^{N-1}
	X_{KN}\right)}~,\label{4.35}\\
  \ln B'^{(N-1)}_J &=& \ln B'_J + \frac{\ln B'_N \ln X_{JN}}
	{\ln \left(A_N A'_N B_N B'_N \prod\limits_{K=1}^{N-1}
	X_{KN}\right)}~, \label{4.36}\\
  \ln X^{(N-1)}_{IJ} &=& \ln X_{IJ} + \frac{\ln X_{IN} \ln X_{JN}}
	{\ln \left(A_N A'_N B_N B'_N \prod\limits_{K=1}^{N-1}
	X_{KN}\right)}~, \label{4.37}
\end{eqnarray}
and ``$\cdots$'' denotes the terms which are independent of the loop
momenta $n^{(J)}$.  They play no essential role in taking the limit
and henceforth we omit those terms.
  Note that the second exponential factor in eq.(\ref{4.32}) takes the
same form as that in eq.(\ref{4.31}) up to a replacement $N\rightarrow
N-1$. Furthermore, completing the squares for $n^{(N-1)}, n^{(N-2)}$,
etc., successively, we obtain,
\begin{eqnarray}
S_o&=& \left( \frac{1}{R_s} \right)^N \sum_{n^{(1)}, \cdots , n^{(N)}}
   \exp  \left[ \alpha'\ln \left(A_N A'_N B_N B'_N \prod_{K=1}^{N-1}
	X_{KN} \right) \right.\label{4.38}   \\
  &&   \times \Bigg( \frac{n^{(N)}}{R_s} -
 \frac{n_2 \ln A_N + n_4 \ln A'_N + n_1 \ln B_N + n_3 \ln B'_N
     +\sum\limits_{K=1}^{N-1} n^{(K)} \ln X_{KN} }
    {R_s \ln \left(A_N A'_N B_N B'_N \prod\limits_{K=1}^{N-1} X_{KN}
        \right)} \Bigg)^2\, \Bigg]\nonumber\\
 &&  \times \exp \left[\alpha'\ln \left(A^{(N-1)}_{N-1}
     A'^{(N-1)}_{N-1} B^{(N-1)}_{N-1} B'^{(N-1)}_{N-1}
	\prod_{K=1}^{N-2}X^{(N-1)}_{KN-1} \right)\right.\nonumber\\
  && \times \Bigg(\frac{n^{(N-1)}}{R_s} -
     \frac{n_2 \ln A^{(N-1)}_{N-1} + \cdots + n_3 \ln B'^{(N-1)}_{N-1}
     + \sum\limits_{K=1}^{N-2} n^{(K)} \ln X^{(N-1)}_{KN-1} }
    {R_s \ln \Bigl(A^{(N-1)}_{N-1} A'^{(N-1)}_{N-1} B^{(N-1)}_{N-1}
           B'^{(N-1)}_{N-1} \prod\limits_{K=1}^{N-2}X^{(N-1)}_{KN}
        \Bigl)} \Bigg)^2 \,\Bigg]\nonumber\\
  && \quad\vdots \nonumber \\
  &&  \times \exp \Biggl[\alpha' \ln \left(A^{(1)}_{1} A'^{(1)}_{1}
	B^{(1)}_{1} B'^{(1)}_{1}\right) \nonumber\\
  &&  \times \left. \left( \frac{n^{(1)}}{R_s} -
     \frac{n_2 \ln A^{(1)}_{1} +n_4 \ln A'^{(1)}_1 +n_1 \ln B^{(1)}_1
     + n_3 \ln B'^{(1)}_{1}}
      {R_s \ln \left(A^{(1)}_{1} A'^{(1)}_{1} B^{(1)}_{1} B'^{(1)}_{1}
         \right) } \right)^2  \right]. \nonumber
\end{eqnarray}
Here, similarly to eqs.(\ref{4.33})$\sim$(\ref{4.37}),
$A_J^{(L)}, A'^{(L)}_J, B_J^{(L)}, B'^{(L)}_J\,(1\leq J\leq L\leq
N-1)$ and $X_{IJ}^{(L)}\,(1\leq I<J\leq L\leq N-1)$ are given by,
\begin{eqnarray}
  \ln A^{(L)}_J&=&\ln A^{(L+1)}_J  + \frac{\ln A^{(L+1)}_{L+1}
	\ln X^{(L+1)}_{JL+1}} {\ln \left(A^{(L+1)}_{L+1}
	A'^{(L+1)}_{L+1} B^{(L+1)}_{L+1} B'^{(L+1)}_{L+1}
	\prod\limits_{K=1}^{L}X^{(L+1)}_{KL+1}\right)}~,\label{4.39}\\
  \ln A'^{(L)}_J&=&\ln A'^{(L+1)}_J  + \frac{\ln A'^{(L+1)}_{L+1}
	\ln X^{(L+1)}_{JL+1}} {\ln \left(A^{(L+1)}_{L+1}
	A'^{(L+1)}_{L+1} B^{(L+1)}_{L+1} B'^{(L+1)}_{L+1}
	\prod\limits_{K=1}^{L}X^{(L+1)}_{KL+1}\right)}~,\label{4.40}\\
  \ln B^{(L)}_J&=&\ln B^{(L+1)}_J  + \frac{\ln B^{(L+1)}_{L+1}
	\ln X^{(L+1)}_{JL+1}} {\ln \left(A^{(L+1)}_{L+1}
	A'^{(L+1)}_{L+1} B^{(L+1)}_{L+1} B'^{(L+1)}_{L+1}
	\prod\limits_{K=1}^{L}X^{(L+1)}_{KL+1}\right)}~,\label{4.41}\\
  \ln B'^{(L)}_J&=&\ln B'^{(L+1)}_J  + \frac{\ln B'^{(L+1)}_{L+1}
	\ln X^{(L+1)}_{JL+1}} {\ln \left(A^{(L+1)}_{L+1}
	A'^{(L+1)}_{L+1} B^{(L+1)}_{L+1} B'^{(L+1)}_{L+1}
	\prod\limits_{K=1}^{L}X^{(L+1)}_{KL+1}\right)}~,\label{4.42}\\
  \ln X^{(L)}_{IJ}&=&\ln X^{(L+1)}_{IJ}  + \frac{\ln X^{(L+1)}_{IL+1}
	\ln X^{(L+1)}_{JL+1}} {\ln \left(A^{(L+1)}_{L+1}
	A'^{(L+1)}_{L+1} B^{(L+1)}_{L+1} B'^{(L+1)}_{L+1}
	\prod\limits_{K=1}^{L}X^{(L+1)}_{KL+1}\right)}~,\label{4.43}
\end{eqnarray}
where $A^{(N)}_J = A_J,\, A'^{(N)}_J = A'_J,\, B^{(N)}_J = B_J,\,
B'^{(N)}_J = B'_J$ and $X_{IJ}^{(N)} = X_{IJ}$.
Similarly to eq.(\ref{3.12}) in section 3, we find that $S_o$ becomes
to have $N$ $\delta$-functions in the L$^3$,
\begin{eqnarray}
 \lim_{\epsilon \rightarrow 0} \,S_o &=& \sum_{n^{(1)}, \cdots , n^{(N)}}
	\prod_{L=1}^N \left( \frac{\pi}{-\alpha'\ln \left(A^{(L)}_{L}
        A'^{(L)}_{L} B^{(L)}_{L} B'^{(L)}_{L}
       \prod\limits_{K=1}^{L-1}X^{(L)}_{KL} \right)}\right)^{\frac{1}{2}}
	 \label{4.44}\\
  &&\times\prod_{L=1}^N
    \delta \left( n^{(L)} -\frac{n_2 \ln A^{(L)}_{L} +
    n_4 \ln A'^{(L)}_{L} +n_1 \ln B^{(L)}_{L} + n_3 \ln B'^{(L)}_{L}
       + \sum\limits_{K=1}^{L-1}n^{(K)} \ln X^{(L)}_{KL} }
    {\ln \left(A^{(L)}_{L} A'^{(L)}_{L} B^{(L)}_{L} B^{(L)}_{L}
       \prod\limits_{K=1}^{L-1}X^{(L)}_{KL}\right) } \right) \nonumber.
\end{eqnarray}
Furthermore, after some calculations using
eqs.(\ref{4.39})$\sim$(\ref{4.43}), eq.(\ref{4.44}) is rewritten in a
more symmetrical form,
\begin{eqnarray}
  \lim_{\epsilon\rightarrow0} \,S_o &=& \sum_{n^{(1)}, \cdots , n^{(N)}}
    \prod_{L=1}^{N}
    \left( \frac{-\pi\ln \left(A^{(L)}_{L} A'^{(L)}_{L} B^{(L)}_{L}
      B'^{(L)}_{L} \prod\limits_{K=1}^{L-1}X^{(L)}_{KL} \right)}
      {\alpha'}\right)^{\frac{1}{2}}\nonumber\\
 &&\times \prod_{L=1}^{N}
    \delta \left( n^{(L)}\ln\left[A_{L}A'_{L}B_{L}B'_{L}
          \prod_{K=1}^{L-1} X_{KL} \prod_{K=L+1}^{N}X_{LK}\right]
	- n_2 \ln A_{L} - n_4 \ln A'_{L} \right.\nonumber\\
 && \hspace{7ex}\left. -n_1 \ln B_{L} - n_3 \ln B'_{L}
       - \sum_{K=1}^{L-1} n^{(K)} \ln X_{KL} - \sum_{K=L+1}^{N}n^{(K)}
	\ln X_{LK} \right).
\end{eqnarray}
Then it is obvious that the condition $n_1=n_2=n_3=n_4$, i.e., all the
external momenta are zero, is necessary in order that the pathological
$\delta(0)$ appears. Hence such a divergence never appears in open
string multi-loop scattering amplitudes.

Next, we consider the L$^3$ in closed string theory. The story is
essentially the same as open string theory, however,
there are two new features in closed string theory, i.e., the existence
of winding modes and the complexification of parameters.

On the $M_{25}\times S^1$, both the external and the loop momenta
along the compact direction can have non-vanishing winding numbers.
For simplicity, however, we assume that the external string states
have vanishing winding numbers \cite{Bil1}. Under this assumption, the
external and the loop momenta along the compact direction are written
by,
\begin{eqnarray}
  p_i^1&=&p_{Ri}^1 = p_{Li}^1 = \frac{n_i}{R_s},\\
  k_{RJ}^1&=&\frac{n^{(J)}}{R_s} - \frac{m^{(J)}R_s}{\alpha'},\\
  k_{LJ}^1&=&\frac{n^{(J)}}{R_s} + \frac{m^{(J)}R_s}{\alpha'},
\end{eqnarray}
where the suffices $R$ and $L$ represent right and left movers of
string states, respectively and $m^{(J)}$ is the winding number of the
string in loop $J$.
Furthermore, on the $M_{25}\times S^1$, the integration over
$k_J^1$ becomes a sum over $n^{(J)}$ and we must also sum up over
$m^{(J)}$. Then the amplitude $I_4^{(N)}$ in (\ref{4.29}) is given by,
\begin{eqnarray}
  I_4^{(N)}&=&\frac{\kappa^{2N+2}}{(4\pi)^{3N+1}}
        \left(\frac{1}{R_s}\right)^N
        \sum_{n^{(1)}, \cdots ,n^{(N)}} \sum_{m^{(1)},\cdots,m^{(N)}}
	\int \prod_{J=1}^{N} d^{25}k'_J
	\int_{|x_j| \leq 1} \prod_{j=1}^{3N+1}d^2x_j\,|G|\nonumber\\
  &&\times |D|^{-\alpha_C(-(p_1-p_3)^2)-2}
           |C|^{-\alpha_C(-(p_2-p_4)^2)-2}
           \prod_{J=1}^N \bigg|A_J A'_J B_J B'_J
	       \prod_{K=J+1}^N X_{JK}\bigg|^{-4}\nonumber \\
  &&  \times   \exp \left[ \frac{\alpha'}{4} \sum_{J=1}^{N} \Biggl\{
         \left((k_{RJ}^1 - p_2^1)^2 +(k'_J-p'_2)^2 \right)\ln A_J
      +  \left((k_{LJ}^1 - p_2^1)^2 +(k'_J-p'_2)^2 \right)\ln \bar{A}_J
          \right. \nonumber\\
  &&\quad+ \left((k_{RJ}^1 - p_4^1)^2 +(k'_J-p'_4)^2 \right)\ln A'_J
        +\left((k_{LJ}^1 - p_4^1)^2 +(k'_J -p'_4)^2 \right)\ln \bar{A'}_J
              \nonumber\\
  &&\quad+\left((k_{RJ}^1 - p_1^1)^2 +(k'_J-p'_1)^2 \right)\ln B_J
         +\left((k_{LJ}^1 - p_1^1)^2 +(k'_J-p'_1)^2 \right)\ln \bar{B}_J
                  \nonumber\\
  &&\quad+\left((k_{RJ}^1 - p_3^1)^2 +(k'_J-p'_3)^2 \right)\ln B'_J
         +\left((k_{LJ}^1 - p_3^1)^2 +(k'_J-p'_3)^2 \right)\ln \bar{B'}_J
                  \nonumber\\
  &&\quad+\sum_{K=J+1}^{N}\left((k_{RJ}^1 - k_{RK}^1)^2 +(k'_J
           -k'_K)^2 \right) \ln X_{JK}\nonumber\\
  &&\quad+ \left.\sum_{K=J+1}^{N}\left((k_{LJ}^1 - k_{LK}^1)^2
	+(k'_J - k'_K)^2\right)\ln \bar{X}_{JK}\Biggr\} \right],
\end{eqnarray}
where $(p'_i)=(p_i^0,p_i^2,p_i^3,\cdots,p_i^{25})$ and
$(k'_J)=(k^0_J,k_J^2,k_J^3,\cdots,k_J^{25})$.
Similarly to eq.(\ref{4.31}), henceforth we only concentrate on the
following $R_s$ dependent part $S_c$ of the amplitude:
\begin{eqnarray}
S_c&=&\left(\frac{1}{R_s} \right)^N \sum_{n^{(1)},\cdots, n^{(N)}}
                                    \sum_{m^{(1)},\cdots, m^{(N)}}
    \exp \left[  \frac{\alpha'}{4} \sum_{J=1}^{N} \left\{
          \left(\frac{n^{(J)}}{R_s}- \frac{m^{(J)}R_s}{\alpha'}
           -\frac{n_2}{R_s} \right)^2 \ln A_J  \right. \right.\nonumber\\
&&    +   \left(\frac{n^{(J)}}{R_s}+ \frac{m^{(J)}R_s}{\alpha'}
               -\frac{n_2}{R_s} \right)^2 \ln \bar{A}_J
      +   \left(\frac{n^{(J)}}{R_s}- \frac{m^{(J)}R_s}{\alpha'}
               -\frac{n_4}{R_s}\right)^2 \ln A'_J \nonumber\\
&&    +   \left(\frac{n^{(J)}}{R_s}+ \frac{m^{(J)}R_s}{\alpha'}
               -\frac{n_4}{R_s}\right)^2 \ln \bar{A'}_J
      +   \left(\frac{n^{(J)}}{R_s}- \frac{m^{(J)}R_s}{\alpha'}
               -\frac{n_1}{R_s}\right)^2 \ln B_J \nonumber \\
&&    +   \left(\frac{n^{(J)}}{R_s}+ \frac{m^{(J)}R_s}{\alpha'}
               -\frac{n_1}{R_s}\right)^2 \ln \bar{B}_J
      +   \left(\frac{n^{(J)}}{R_s}- \frac{m^{(J)}R_s}{\alpha'}
               -\frac{n_3}{R_s}\right)^2 \ln B'_J\nonumber \\
&&    +   \left(\frac{n^{(J)}}{R_s}+ \frac{m^{(J)}R_s}{\alpha'}
               -\frac{n_3}{R_s}\right)^2 \ln \bar{B'}_J\nonumber\\
&&    + \sum_{K=J+1}^{N}
         \left( \left(\frac{n^{(J)}}{R_s}- \frac{m^{(J)}R_s}{\alpha'}
               -\frac{n^{(K)}}{R_s}
               +\frac{m^{(K)}R_s}{\alpha'}  \right)^2  \ln X_{JK}\right.
               \nonumber\\
&&\hspace{10ex}+ \left. \left. \left.
           \left (\frac{n^{(J)}}{R_s}+ \frac{m^{(J)}R_s}{\alpha'}
             -\frac{n^{(K)}}{R_s}-\frac{m^{(K)}R_s}{\alpha'}\right)^2
	     \ln \bar{X}_{JK}\right) \right\} \right] \nonumber\\
&=&  \left(\frac{1}{R_s} \right)^N \sum_{n^{(1)},\cdots, n^{(N)}}
      \exp \left[  \frac{\alpha'}{2} \sum_{J=1}^{N} \left\{
         \left(\frac{n^{(J)}}{R_s} -\frac{n_2}{R_s} \right)^2
         \mbox{Re}\left(\ln A_J \right)  \right. \right.\nonumber \\
&&    +  \left(\frac{n^{(J)}}{R_s} -\frac{n_4}{R_s} \right)^2
         \mbox{Re}\left(\ln A' \right)
      +  \left(\frac{n^{(J)}}{R_s} -\frac{n_1}{R_s} \right)^2
         \mbox{Re}\left(\ln B_J \right) \nonumber \\
&&    +  \left(\frac{n^{(J)}}{R_s} -\frac{n_3}{R_s} \right)^2
         \mbox{Re}\left(\ln B'_J \right)
     +  \sum_{K=J+1}^{N} \left.\left.
         \left(\frac{n^{(J)}}{R_s}-\frac{n^{(K)}}{R_s}\right)^2
         \mbox{Re}\left(\ln X_{JK} \right) \right\}\right]
         \tilde{S_c}  \label{eqScl}\,,
\end{eqnarray}
where,
\begin{eqnarray}
  \tilde{S_c}&=& \sum_{m^{(1)},\cdots, m^{(N)}} \exp
	\left[ -\pi\, \vec{m}^T M^{-1} \vec{m} + 2\pi i
	\,\vec{m}^T \vec{x}\right],\\
  M^{-1}&=&\frac{R_s^2}{2\pi \alpha'}\,
    \left[\begin{array}{lllll}
	  a_{11}& a_{12} & & \cdots     &a_{1N}\\
        	& a_{22} & a_{23}&\cdots&a_{2N}\\
	        &        & \ddots&      &\vdots\\
	        &\hbox{\Large 0}&  &\ddots& a_{N-1N}\\
		&        &       &      & a_{NN}
	\end{array}\right],\label{4.52}\\
  \vec{m}&=& \left[\begin{array}{c}
	m^{(1)} \\  m^{(2)} \\  \vdots  \\  m^{(N)} \\
		\end{array}\right],\hspace{5ex}
	\vec{x}= \left[\begin{array}{c}
		  x^{(1)} \\  x^{(2)} \\  \vdots  \\  x^{(N)} \\
		\end{array}\right],\\
  a_{JJ}&=& -\mbox{Re} \left\{\ln \left( A_J A'_J B_J B'_J
           \prod_{K=1}^{J-1}X_{KJ}\prod_{K=J+1}^{N}X_{JK}\right)\right\},\\
  a_{JK}&=& 2\,\mbox{Re}\,(\ln X_{JK}),\quad a_{KJ}=0, \hspace{5ex}(J<K),\\
  x^{(J)}&=&-\frac{1}{2\pi}\Biggl[
	n^{(J)} \mbox{Im} \left\{\ln \left( A_{J} A'_{J} B_{J} B'_{J}
        \prod_{K=1}^{J-1} X_{KJ} \prod_{K=J+1}^{N}X_{JK}\right)\right\}
	-n_2 \mbox{Im} (\ln A_{J})
	\nonumber\\
  && - n_4 \mbox{Im}(\ln A'_{J}) -n_1 \mbox{Im} (\ln B_{J})
	- n_3 \mbox{Im}(\ln B'_{J}) \nonumber\\
  && - \sum_{K=1}^{J-1} n^{(K)}  \mbox{Im}(\ln X_{KJ})
       - \sum_{K=J+1}^{N} n^{(K)}  \mbox{Im}(\ln X_{JK})\Biggr].
\end{eqnarray}
Note that the (first) exponential factor in eq.(\ref{eqScl})
takes the same form as that in eq.(\ref{4.31}) by replacing
$\alpha' \rightarrow \frac{\alpha'}{2}$ and $\ln  \rightarrow
\mbox{Re}\ln$.
Hence, the remaining problem is to take the L$^3$ of
$\tilde{S_c}$. This can be solved by using the Poisson resummation
formula \cite{Bil1}:
\begin{equation}
  \sum_{m^{(1)},\cdots, m^{(N)}} \exp
       \left[ -\pi \vec{m}^T M^{-1} \vec{m} + 2\pi i \vec{m}^T
        \vec{x}\right]
  =  \sum_{m^{(1)},\cdots, m^{(N)}}(\det M)^{\frac{1}{2}}\,
     \exp \left[-\pi(\vec{m}+\vec{x})^{T} M (\vec{m} +\vec{x})\right].
\end{equation}
Note that the $N\times N$ matrix $M$ is proportional
to $\frac{1}{R_s^2}$  due to eq.(\ref{4.52}) and the $(\det
M)^{\frac{1}{2}}$ on the r.h.s. gives $(\frac{1}{R_s})^N$ factor.
Hence, after some calculations, we also obtain $N$ $\delta$-functions
in the L$^3$:
\begin{eqnarray}
 \lim_{\epsilon \rightarrow 0} \,\tilde{S_c}&=&
	(2\pi)^N\sum_{m^{(1)}, \cdots , m^{(N)}}
	\prod_{L=1}^{N} \delta \left( n^{(L)} \mbox{Im}\left\{
            \ln \left( A_{L} A'_{L} B_{L} B'_{L}
            \prod_{K=1}^{L-1} X_{KL} \prod_{K=L+1}^{N}X_{LK}
            \right)\right\} \right.\nonumber\\
  && \quad- n_2 \mbox{Im}(\ln A_{L}) - n_4 \mbox{Im}(\ln A'_{L})
     - n_1 \mbox{Im}(\ln B_{L}) - n_3 \mbox{Im}(\ln B'_{L})\nonumber\\
  &&\quad\left. - \sum_{K=1}^{L-1} n^{(K)} \mbox{Im}(\ln X_{KL})
        - \sum_{K=L+1}^{N} n^{(K)} \mbox{Im}(\ln X_{LK})
          - 2 \pi  m^{(L)}\right).
\end{eqnarray}
Thus, putting those together, we find that the $S_c$ gives $N$ complex
$\delta$-functions in the L$^3$,
\begin{eqnarray}
 \lim_{\epsilon \rightarrow 0} \,S_c&=& (2 \pi)^N\sum_{n^{(1)},
	\cdots, n^{(N)}}\sum_{m^{(1)}, \cdots , m^{(N)}}
	\prod\limits_{L=1}^{N} \left( \frac{-2\pi \mbox{Re}
           \ln \left(A^{(L)}_{L} A'^{(L)}_{L} B^{(L)}_{L}
            B'^{(L)}_{L} \prod\limits_{K=1}^{L-1}X^{(L)}_{KL} \right)}
           {\alpha'}\right)^{\frac{1}{2}}\nonumber\\
  &&\times \prod_{L=1}^{N} \delta^{(2)} \left( n^{(L)}
	\ln \left( A_{L} A'_{L} B_{L} B'_{L} \prod_{K=1}^{L-1} X_{KL}
	 \prod_{K=L+1}^{N}X_{LK}\right) - n_2 \ln A_{L}
	\right.\nonumber\\
  &&\quad - n_4 \ln A'_{L}   -n_1 \ln B_{L} - n_3 \ln B'_{L} \nonumber\\
  && \quad\left.  - \sum_{K=1}^{L-1} n^{(K)} \ln X_{KL}
       - \sum_{K=L+1}^{N}n^{(K)} \ln X_{LK}
                        - 2 \pi i m^{(L)}\right).
\end{eqnarray}
  From the above equation, it is also obvious that $n_1=n_2=n_3=n_4$ is
a necessary condition for an appearance of the pathological $\delta(0)$.
Hence such a divergence does not occur in closed string multi-loop
scattering amplitudes.

We have examined the L$^3$ of multi-loop open and closed string
scattering amplitudes. In conclusion, {\em since by construction these
amplitudes consist of all the possible propagators in the field
theoretical Feynman diagrams, which is the root of s-t channel
duality, we can take the L$^3$ of these scattering amplitudes, i.e.,
the zero-mode loop divergences do not appear.} Note that it is obvious
that the nonexistence of the $\delta(0)$ is not essentially due to the
existence of the winding modes (cf. \cite{Bil1}).
%%%%%%%%%%%%%%%%%%%%%%%%%%%%%%%%%%%%%%%%%%%%%%%%%%%%%%%%%%%%%%%%%%%%
\section{The L$^3$ of vacuum amplitudes in string theory}
%%%%%%%%%%%%%%%%%%%%%%%%%%%%%%%%%%%%%%%%%%%%%%%%%%%%%%%%%%%%%%%%%%%%
In this section, we discuss the L$^3$ of vacuum amplitudes in bosonic
string theory. For simplicity, we restrict ourselves to bosonic closed
string theory.

	In the previous section, we have seen that the factor
$\frac{1}{R_s^2}\sim\frac{1}{\epsilon^2}$ has come out for each loop
from the integration measure of the compactified loop momentum and the
Poisson resummation of the winding modes.
This factor has combined with an appropriate exponential function
$\exp\left(-\left(\frac{\pi}{R_s^2}\right)\vert\cdots\vert^2\right)$
to give a complex $\delta$-function for each loop. We have shown that
this complex $\delta$-function can become $\delta^{(2)}(0)$ iff all
the external momenta in the compact direction $P_i^1=\frac{N_i}{R_s}$
are zero at multi-loop order.
Thus, this pathological $\delta^{(2)}(0)\sim\frac{1}{R_s^2}$ does not
appear in the multi-loop string scattering amplitudes. As for the
vacuum amplitudes which have no external lines, however, it is obvious
that this pathological situation inevitably occurs, i.e., the zero-mode
loop divergences always appear.
In fact, as is well known, in a T-dual theory this
$\delta^{(2)}(0)\sim \frac{1}{R_s^2}$ for each loop is absorbed in the
T-dual coupling constant $\tilde{\kappa}=e^{\tilde{\phi}}$,
\begin{equation}
	\tilde{\kappa}^2=\kappa^2\frac{\alpha'}{R_s^2},
\end{equation}
where $\kappa$ is the original closed string coupling constant.
In the L$^3$, since we shall keep the original string coupling
constant fixed, this dual coupling constant diverges. Thus the
zero-modes become strongly coupled and the cosmological constant can
not be determined perturbatively.
In particular, since the closed string theory includes gravity, we
shall take this problem seriously. Of course, we should notice that
the M-theory or superstring does not suffer from this problem due to
supersymmetry, and neither does the direct DLCQ theory due to the
triviality of the vacuum.
%%%%%%%%%%%%%%%%%%%%%%%%%%%%%%%%%%%%%%%%%%%%%%%%%%%%%%%%%%%%%%%%%%%%
\section{Conclusion and Discussion}
%%%%%%%%%%%%%%%%%%%%%%%%%%%%%%%%%%%%%%%%%%%%%%%%%%%%%%%%%%%%%%%%%%%%
In this paper, we have argued whether perturbative string multi-loop
amplitudes have well-defined light-like limit or not.
To answer this question, we have used the multi-loop string
scattering amplitudes constructed by the method of \cite{KSV}.
In this method, since we can keep the integral, or sum, over the loop
momenta, we have shown that the perturbative string multi-loop
scattering amplitudes have well-defined light-like limit,
which was conjectured in ref.\cite{Bil2}. This result depends on the
striking stringy nature. In fact, we must add all possible Feynman
diagrams in a certain order of perturbation theory and the zero-mode
loop diagrams among them diverge as $\delta(0)$ in field theory, while
in string theory, since a scattering amplitude in a certain order
of perturbation contains the resonances in the all possible field
theoretical Feynman diagrams in the same order of perturbation, the
zero-mode loop divergences do not appear.

We also have discussed the light-like limit of the vacuum amplitudes
in bosonic string theory. These amplitudes obviously diverge as
$\delta(0)$ even in string theory. Therefore, we need supersymmetry
for this limit to be well-defined.
It seems that this result prevents us from extracting the information
of supersymmetry breaking in M-theory or superstring from the finite
$N$ Matrix theory. In fact, it seems that the existence of
non-supersymmetric Matrix string theory is questionable \cite{BM}.
%%%%%%%%%%%%%%%%%%%%%%%%%%%%%%%%%%%%%%%%%%%%%%%%%%%%%%%%%%%%%%%%%%%%
\appendix
%%%%%%%%%%%%%%%%%%%%%%%%%%%%%%%%%%%%%%%%%%%%%%%%%%%%%%%%%%%%%%%%%%%%
\section{KSV construction of $N$-point tree amplitude}
%%%%%%%%%%%%%%%%%%%%%%%%%%%%%%%%%%%%%%%%%%%%%%%%%%%%%%%%%%%%%%%%%%%%
In this appendix, we calculate the tachyon $N$-point tree amplitude in
bosonic open string theory by the KSV method \cite{KSV}.
We shall determine the form of the function $y=f(x;a_1,a_2,a_3,a_4)$,
i.e., (\ref{4.17})$\sim$(\ref{4.19}), by comparing the eight-point
amplitude by the KSV method with that in the operator formalism.
Then by using this function, we will show that the $N$-point tree
amplitude constructed by the KSV method agrees with the one in the
operator formalism.
Similarly to the four-point amplitude in section \ref{secMLST}, we
consider the dual diagrams to the field theoretical Feynman diagrams
connected by s-t channel duality for the $N$-point tree amplitude.
That is, we consider the $N$-polygons defined by $N$ sides which
are dual to the $N$ external lines.
Each diagonal of the $N$-polygon is dual to a propagator.
In each dual diagram, we always have $N-3$ diagonals which correspond
to the $N-3$ propagators in the field theoretical Feynman diagram.

\begin{figure}[htbp]
\centerline{\epsfbox{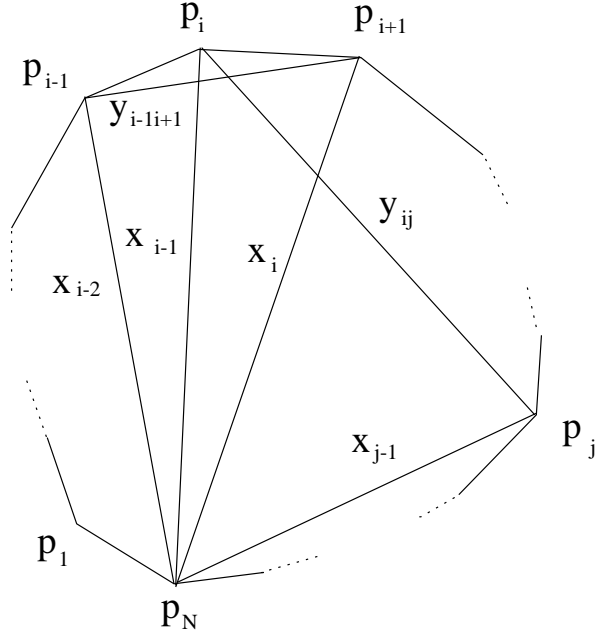}}
\caption{(Dual) diagram for the $N$-point tree amplitude.}
\label{figDDNT}
\end{figure}
We define the dual external momenta $p_i$ ($i=1,\cdots,N$) as
follows (see (\ref{3.2}) $\sim $ (\ref{3.5})):
\begin{equation}
    \begin{array}{lll}
	P_1&=& p_1 - p_N ,\\
	P_2&=& p_2 - p_1 ,\\
	&&\ldots \\
	P_N&=& p_N - p_{N-1},
    \end{array}\label{eqDM}
\end{equation}
and we assign $p_i$ ($i =1,\cdots,N$) to the $N$ vertices of the
$N$-polygon, respectively (Figure \ref{figDDNT}).
The $\frac{N(N-3)}{2}$ lines connecting the
vertices $p_{i-1}$ and $p_j$  ($2\leq i<j\leq N-1,\, 3\leq i<j=N$) are
denoted by $y_{i-1\,j}$ and we choose $N-3$ independent lines
$x_i\equiv y_{i+1N}$ ($i=1,\cdots,N-3$) among them (Figure \ref{figDDNT}).
Then we can write down the $N$-point tree amplitude by the KSV method:
\begin{equation}
  A_{N}^{(0)} = g^{N-2}\int _0^1 \prod_{i=1}^{N-3} dx_i\,
	\rho^{(0)}(x_i)
	\prod_{i=1}^{N-3}x_i^{-\alpha(-(p_{i+1}-p_N )^2)-1} \prod_{2
	\leq i <j \leq N-1 } y_{i-1j}^{-\alpha(-(p_{i-1}-p_j
	)^2)-1}\label{eqNKSV},
\end{equation}
where $\rho^{(0)}(x_i)$ is determined by the requirement that
the integral volume element maintains the crossing
symmetry\cite{KaN1,KaN2},
\begin{equation}
    \rho^{(0)}(x_i)=
	\frac{1}{\prod\limits_{2\leq i<N-2}(1-x_ix_{i-1})}.
\end{equation}

Next we shall determine the forms of the functions $y_{i-1j}(x_k)\,
(2\leq i<j\leq N-1)$  and compare (\ref{eqNKSV}) with the tree amplitude
in the operator formalism. In the operator formalism the $N$-point
tree amplitude is given by \cite{Sch},
\begin{equation}
  A_{Nop}^{(0)} = g^{N-2}\int_0^1 \prod_{i=1}^{N-3} dx_i\,
	\prod_{i=1}^{N-3}x_i^{-\alpha({s_i} )-1}
	\prod_{2 \leq i <j \leq N-1 }
	(1-x_{ij})^{2\alpha' (P_i \cdot  P_j)},\label{eqNTO}
\end{equation}
where
\begin{eqnarray}
	x_{ij}&=& x_{i-1}x_i\cdots x_{j-2},\\
	s_i&=& - (P_1 + \ldots +P_{i+1})^2,
\end{eqnarray}
  From eqs.(\ref{eqDM}) we obtain,
\begin{eqnarray}
	(p_{i-1} -p_j)^2 &=& (P_i + \ldots +P_j)^2,\\
	s_i &=& -(p_{i+1} -p_N)^2.
\end{eqnarray}
Then, after some calculations, (\ref{eqNTO}) is rewritten by,
\begin{eqnarray}
  A_{Nop}^{(0)}&=& g^{N-2}\int_0^1 \prod_{i=1}^{N-3} dx_i
	\rho^{(0)} (x_i)
	\prod_{i=1}^{N-3}x_i^{-\alpha(-(p_{i+1}-p_N)^2)-1}
	\nonumber\\
  &&\times \prod_{2\leq i <j\leq N-1}
	\left(\frac{1-x_{ij}}{1-x_{i-1j}}\,
	\frac{1-x_{i-1j+1}}{1-x_{ij+1}}
	\right)^{-\alpha (-(p_{i-1}-p_j)^2)-1}\label{eqNTO2}.
\end{eqnarray}
Here, we have used the equations which $x_0= x_{N-2}=0$, i.e.,
$x_{1j} = x_{iN}=0$.
 
\begin{figure}[htbp]
\centerline{\epsfbox{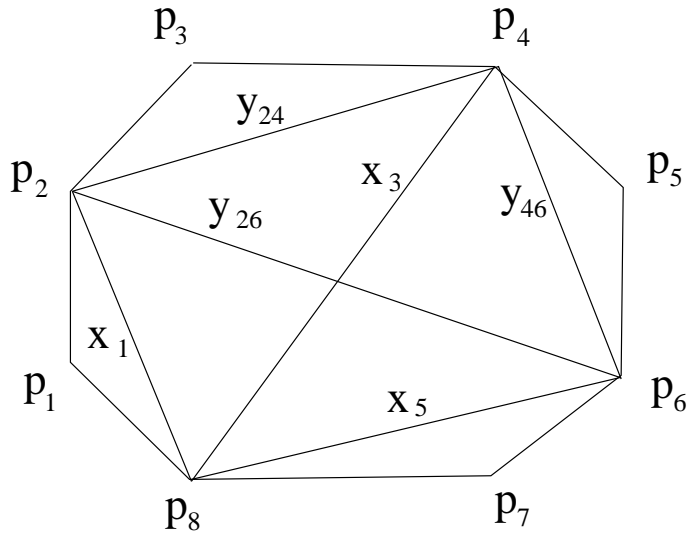}}
\caption{(Dual) diagram for the eight-point tree amplitude.}
\label{figET}
\end{figure}

Now we concentrate on the $N=8$ case and determine the form
of the function (\ref{4.9}), $y=f(x;a_1,a_2,a_3,a_4)$. We can easily
see that (\ref{eqNKSV}) agrees with (\ref{eqNTO2}) if
\begin{equation}
  y_{i-1j}= \frac{1-x_{ij}}{1-x_{i-1j}}\,
	\frac{1-x_{i-1j+1}}{1-x_{ij+1}}\label{eqGY}.
\end{equation}
For example, the lines $y_{26}$, $y_{24}$ and $y_{46}$ are written by,
\begin{eqnarray}
  y_{26}&=&\frac{1-x_2x_3x_4}{1-x_1x_2x_3x_4}\,
	\frac{1-x_1x_2x_3x_4x_5}{1-x_2x_3x_4x_5},\\
  y_{24}&=&\frac{1-x_2}{1-x_1x_2}\,\frac{1-x_1x_2x_3}{1-x_2x_3},\\
  y_{46}&=&\frac{1-x_4}{1-x_3x_4}\,\frac{1-x_3x_4x_5}{1-x_4x_5}.
\end{eqnarray}
In Figure \ref{figET} we see that the line $y_{26}$ is a diagonal of
the quadrilateral whose sides are ($x_1,y_{24}, y_{46},x_5$) and the
other diagonal of this quadrilateral is $x_3$. And hence we can write,
\begin{equation}
	y_{26}= f(x_3 ; x_1,y_{24},y_{46}, x_5).
\end{equation}
Then we have obtained eqs.(\ref{4.17})$\sim$(\ref{4.19}).
Note that when $N < 8$, we cannot draw a quadrilateral whose all sides
are diagonals of the $N$-polygon of the dual diagram.

Next by using these eqs.(\ref{4.17})$\sim$(\ref{4.19}), we show
that (\ref{eqGY}) holds for an arbitrary $N$, i.e., the $N$-point tree
amplitude by the KSV method agrees with that in the operator formalism.
Let us consider the lines $y_{i-1i+1}\ (2\leq i\leq N-2)$ in
Figure \ref{figDDNT}.
Since $y_{i-1i+1}$ is a diagonal of the quadrilateral
whose sides are ($x_{i-2},0,0,x_{i}$) with the other diagonal $x_{i-1}$,
we have
\begin{eqnarray}
  y_{i-1i+1}&=& f(x_{i-1}; x_{i-2},0,0,x_i)\nonumber\\
	&=& \frac{1-x_{i-1}}{1-x_{i-1}x_{i-2}}
	\frac{1-x_{i-2}x_{i-1}x_i}{1-x_{i-1}x_{i}}.
\end{eqnarray}
Then we will prove (\ref{eqGY}) by mathematical induction \cite{KSV}.
Suppose the equation holds for $y_{i-1 j}$.
Since the line $y_{i-1j+1}$ is a diagonal of the quadrilateral
whose sides are ($x_{i-2},y_{i-1j},0,x_j$), we have
\begin{eqnarray}
  y_{i-1j+1} &=& f (x_{j-1}; x_{i-2},y_{i-1j},0,x_j)\nonumber\\
  &=& \frac{1-x_{j-1} \alpha}{1-x_{j-1}\alpha x_{i-2}}
	\frac{1-x_{j-1}\alpha x_{i-2}x_j}{1-x_{j-1}\alpha
	  x_{j}}\label{eqGY2}.
\end{eqnarray}
where $\alpha$ is defined implicitly by,
\begin{equation}
  y_{i-1j} = \frac{1-\alpha}{1-\alpha x_{i-2}}\,
	\frac{1-\alpha x_{j-1} x_{i-2}}{1-\alpha x_{j-1}}\,.
\end{equation}
Here by the assumption of mathematical induction, we obtain,
$\alpha = x_{ij}=x_{i-1} \cdots x_{j-2}$.
Plugging this into (\ref{eqGY2}),
we obtain $y_{i-1j+1}=\frac{1-x_{ij+1}}{1-x_{i-1j+1}}\,
\frac{1-x_{i-1j+2}}{1-x_{ij+2}}$. Hence we have proven (\ref{eqGY})
for an arbitrary $N$.
%%%%%%%%%%%%%%%%%%%%%%%%%%%%%%%%%%%%%%%%%%%%%%%%%%%%%%%%%%%%%%
\section{KSV construction of $N$-point one-loop amplitude}
%%%%%%%%%%%%%%%%%%%%%%%%%%%%%%%%%%%%%%%%%%%%%%%%%%%%%%%%%%%%%%
In this appendix we construct the tachyon $N$-point one-loop amplitude
in bosonic open string theory by the KSV method \cite{KSV}.
Each dual diagram to the field theoretical Feynman diagram is a
$N$-polygon with an internal point and $N$ internal lines which
connect either two of the vertices or a vertex and an internal point
properly and do not intersect each other. The $N$ sides of the polygon
and the internal point are dual to the $N$ external lines and the
loop in the Feynman diagram, respectively.
In order to construct the amplitude we shall consider all
topologically inequivalent lines connecting either two of the vertices
or a vertex and the internal point of the $N$-polygon.
As in the previous appendix, we assign the dual momenta $p_i\,
(i=1,\cdots,N)\, (\ref{eqDM})$ to the $N$ vertices of the $N$-polygon
and we call the internal point $o_1$.
The $N$ lines connecting the vertices $p_i$ and the internal point
$o_1$, which do not exist in the tree case, are denoted by $x_i$.
As for the lines connecting two vertices of the $N$-polygon, a pair of
vertices do not uniquely determine a line in this case due to the
existence of the point $o_1$ and the key to constructing the one-loop
amplitude is to completely determine those lines.

\begin{figure}
	\epsfxsize80mm
	\centerline{\epsfbox{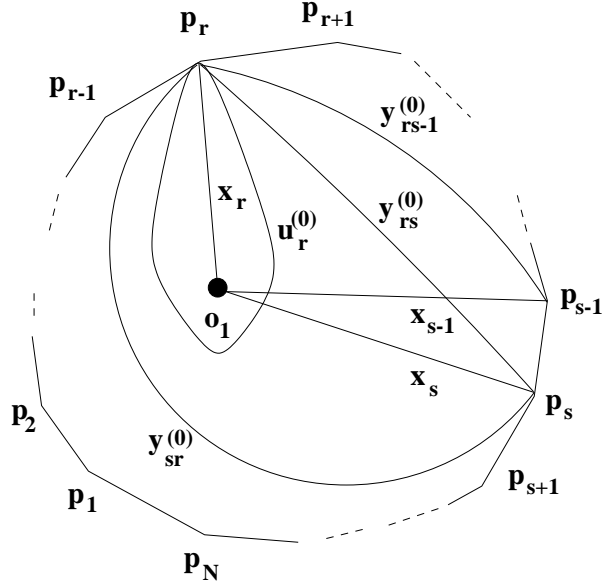}}
	\caption{(Dual) diagram for the $N$-point one-loop amplitude.}
	\label{figNOL}
\end{figure}

	From $p_r$ to $p_s$ we draw a line clockwise with the point
$o_1$ kept to the right.  So we can draw two lines for each pair of
the vertices $p_r$ and $p_s$. One is from $p_r$ to $p_s$ and the other
is from $p_s$ to $p_r$. They are denoted by $y^{(0)}_{rs}$ and
$y^{(0)}_{sr}$, respectively (see Figure \ref{figNOL}). Note that
without the point $o_1$, i.e., for the tree amplitude, $y^{(0)}_{rs}$
and $y^{(0)}_{sr}$ are (topologically) equivalent, but this is not the
case with $o_1$.
We should notice that $y^{(0)}_{r r+1}$ is a side of the polygon and
hence $y^{(0)}_{r r+1}=0$, but $y^{(0)}_{r+1 r}$ is not.
Furthermore we can draw a non-trivial line from $p_r$ going around the
point $o_1$ and coming back to $p_r$, which is denoted by $u^{(0)}_r$
(Figure \ref{figNOL}).

	As in the tree case, $x_i$, $y^{(0)}_{rs}, y^{(0)}_{sr},
y^{(0)}_{r+1 r}$ and $u^{(0)}_r$ are not independent and we can take
$x_i \, (i=1,\cdots,N)$ as independent parameters and determine
$y^{(0)}_{rs}, y^{(0)}_{sr}, y^{(0)}_{r+1 r} $ and $u^{(0)}_r$ as
functions of $x_i$.
  We explain how to obtain $y^{(0)}_{rs} = y^{(0)}_{rs}(x_i)\,
(1\leq r<s-1<N\,,\,(r,s)\neq(1,N))$ below.
We shall see that $y^{(0)}_{rs}$ is a diagonal
of a quadrilateral $o_1p_rp_{s-1}p_s$ whose sides are
$(x_r,y^{(0)}_{rs-1},0,x_s)$ and hence we have,
\begin{equation}
	y^{(0)}_{rs}= f(x_{s-1}; x_r,y^{(0)}_{rs-1},0,x_s)\,.
\end{equation}
where $f$ is defined by (\ref{4.17})$\sim$(\ref{4.19}).
Similarly $y^{(0)}_{rs-1}$ is a diagonal of a quadrilateral
$o_1p_rp_{s-2}p_{s-1}$, $y^{(0)}_{rs-2}$ is a diagonal of a quadrilateral
$o_1p_rp_{s-3}p_{s-2}$ and so on. Then we have the following
series of equations:
\begin{eqnarray}
  y^{(0)}_{rs-1}&=&f(x_{s-2}; x_r,y^{(0)}_{rs-2},0,x_{s-1})\,,\\
  y^{(0)}_{rs-2}&=&f(x_{s-3}; x_r,y^{(0)}_{rs-3},0,x_{s-2})\,,\\
  &&\cdots\nonumber\\
  y^{(0)}_{rr+4}&=&f(x_{r+3}; x_r,y^{(0)}_{rr+3},0,x_{r+4})\,,\label{eqY4}\\
  y^{(0)}_{rr+3}&=&f(x_{r+2}; x_r,y^{(0)}_{rr+2},0,x_{r+3})\,,\label{eqY}\\
  y^{(0)}_{rr+2}&=&f(x_{r+1}; x_r,0,0,x_{r+2})\,\label{eqY1}.
\end{eqnarray}
Eq.(\ref{eqY}) gives,
\begin{eqnarray}
  y^{(0)}_{rr+3}&=&
	\frac{1-x_{r+2}\alpha}{1-x_{r+2}\alpha x_r}\,
	\frac{1-x_{r+2}\alpha x_rx_{r+3}}{1-x_{r+2}\alpha
	x_{r+3}},\\
  y^{(0)}_{rr+2}&=&\frac{1-\alpha}{1-\alpha x_r}\,
	\frac{1-\alpha x_rx_{r+2}}{1-\alpha x_{r+2}}\,,\label{eqY2A}
\end{eqnarray}
while (\ref{eqY1}) is rewritten by,
\begin{equation}
  y^{(0)}_{rr+2}=\frac{1-x_{r+1}}{1-x_{r+1}x_r}\,
	\frac{1-x_{r+1}x_r x_{r+2}}{1-x_{r+1}x_{r+2}}\label{eqY2}.
\end{equation}
Eqs.(\ref{eqY2A}) and (\ref{eqY2}) lead to $\alpha = x_{r+1}$ and
hence we obtain,
\begin{equation}
  y^{(0)}_{rr+3}=\frac{1-x_{r+1}x_{r+2}}{1-x_rx_{r+1}x_{r+2}}\,
	\frac{1-x_rx_{r+1}x_{r+2}x_{r+3}}{1-x_{r+1}x_{r+2}x_{r+3}}\,.
	\label{eqY3}
\end{equation}
Similarly we can determine $y^{(0)}_{rr+4}$ by eqs.(\ref{eqY4}) and
(\ref{eqY3}) and so on. Then we obtain,
\begin{equation}
  y^{(0)}_{rs}=	\frac{1-c_{s-1\,r}}{1-c_{sr}}\,
	\frac{1-c_{s\,r-1}}{1-c_{s-1\,r-1}}\,,
         \quad (1\leq r<s-1<N\,,\,(r,s) \neq (1,N))\,,
\end{equation}
where
\begin{equation}
	c_{sr} = x_{r+1}x_{r+2}\cdots x_{s-1}x_s\,.
\end{equation}

In a similar way, we get,
\begin{eqnarray}
  y^{(0)}_{sr}&=&f(x_{r-1}; x_s,y^{(0)}_{sr-1},0,x_r)
         \nonumber\\
  &=&	\frac{1-(w/c_{s-1\,r})}{1-(w/c_{sr})}\,
	\frac{1-(w/c_{s\,r-1})}{1-(w/c_{s-1\,r-1})},
	~~(1\leq r<s-1<N\,,\,(r,s) \neq (1,N)),\label{eqYsr}
\end{eqnarray}
where
\begin{equation}
	w = \prod_{i=1}^{N} x_i\,.
\end{equation}
We should notice that $y^{(0)}_{r+1\,r}$ can be regarded as a diagonal
of the quadrilateral $p_{r+1}p_{r-1}p_ro_1$ whose sides are
$(y^{(0)}_{r+1\,r-1},0,x_r,x_{r+1})$ \cite{KSV} and hence we obtain,
\begin{equation}
	y^{(0)}_{r+1\,r}=f(x_{r-1};x_{r+1},y^{(0)}_{r+1\,r-1},0,x_r)\,.
	\label{eqYr1}
\end{equation}
Then eqs.(\ref{eqYsr}) and (\ref{eqYr1}) lead to
\begin{eqnarray}
  y^{(0)}_{r+1\,r}&=&
	\frac{1-w}{1-(w/c_{r+1r})}\,
	\frac{1-(w/c_{r+1\,r-1})}{1-(w/c_{r\,r-1})}\nonumber\\
  &=&	\frac{1-w}{1-(w/x_{r+1})}\,
	\frac{1-(w/x_rx_{r+1})}{1-(w/x_r)}\,,
	\quad (0\leq r<N,\,x_0\equiv x_N).
	\label{eqYr}
\end{eqnarray}
Furthermore $u^{(0)}_{r+1}$ can be regarded as a diagonal
of the ``quadrilateral'' $p_{r+1}p_rp_{r+1}o_1$ whose sides are
$(y^{(0)}_{r+1\,r},0,x_{r+1},x_{r+1})$ \cite{KSV} and hence we obtain,
\begin{equation}
	u^{(0)}_{r+1}= f(x_r;x_{r+1},y^{(0)}_{r+1\,r},0,x_{r+1})\,.
	\label{eqYu1}
\end{equation}
Then eqs.(\ref{eqYu1}) and (\ref{eqYr}) lead to
\begin{equation}
  u^{(0)}_r= \frac{(1-x_rw)(1-(w/x_r))}{(1-w)^2}\,,
	\quad (1\leq r \leq N).\label{eqU0}
\end{equation}

\begin{figure}
\epsfxsize140mm
\centerline{\epsfbox{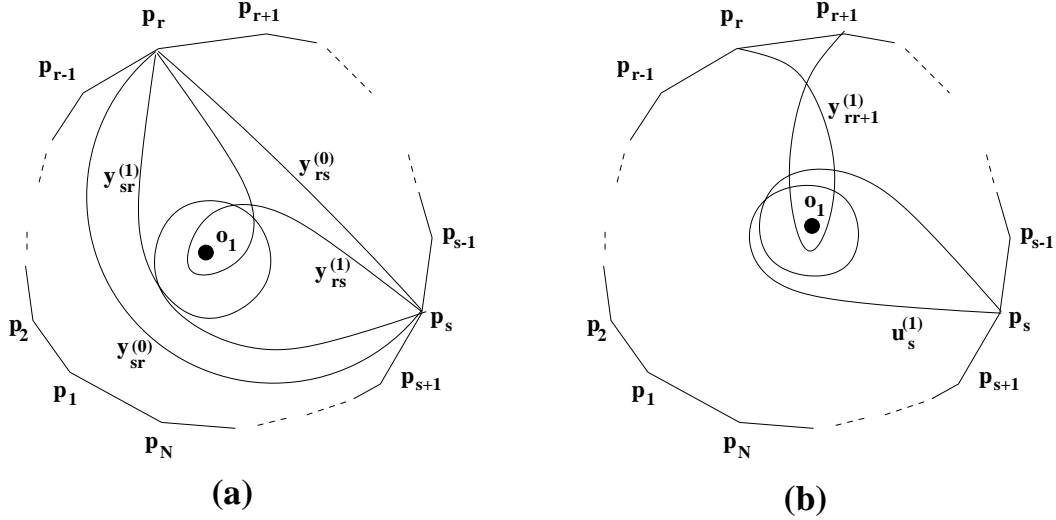}}
\caption{Winding lines in (dual) diagram.}
\label{figWL}
\end{figure}

	We have seen that there are two topologically inequivalent
lines, $y^{(0)}_{rs} $and $y^{(0)}_{rs}$, connecting a pair of
vertices of the $N$-polygon in general. In fact there are infinite
number of lines. In Figure \ref{figWL}(a) we find that a line
$y^{(1)}_{rs}$ which connects two vertices,  $p_r$ and $p_s$, with
winding the point $o_1$ is topologically inequivalent to
$y^{(0)}_{rs}$.  Note that $y^{(1)}_{rs}$ is drawn clockwise from
$p_r$ to $p_s$ with the point $o_1$ always kept on the right and hence
$y^{(1)}_{rs}\neq y^{(1)}_{sr}$.
	Then we have infinite number of inequivalent lines
$y^{(n)}_{rs}, y^{(n)}_{sr}\, (n=0,1,2,\cdots)$ connecting a pair of
vertices, $p_r$ and $p_s$ ($1\leq r<s-1<N$).
	We know that $y^{(0)}_{rs}$ is a diagonal
of a quadrilateral $o_1p_rp_{s-1}p_s$ whose sides are
$(x_r,y^{(0)}_{rs-1},0,x_s)$ and we may regard all the $y^{(n)}_{rs}$
as diagonals of the same quadrilateral $o_1p_rp_{s-1}p_s$.
	However, since  $y^{(n)}_{rs}$ crosses all the $x_i$ $n$ times,
we regard the other diagonal of the quadrilateral as $x_{s-1}w^n$ and
hence we get ($n=0,1,2,\cdots$),
\begin{eqnarray}
  y^{(n)}_{rs}&=& f(x_{s-1}w^n; x_r,y^{(0)}_{rs-1},0,x_s)\nonumber\\
  &=& \frac{1-c_{s-1\,r}w^n}{1-c_{sr}w^n}\,
    \frac{1-c_{s\,r-1}w^n}{1-c_{s-1\,r-1}w^n},
    \hspace{3ex}(1\leq r<s-1<N\,,\,(r,s) \neq (1,N)).\label{eqYnrs}
\end{eqnarray}
Similarly we obtain ($n=0,1,2,\cdots$),
\begin{eqnarray}
  y^{(n)}_{sr}&=&f(x_{r-1}w^n; x_s,y^{(0)}_{sr-1},0,x_r)\nonumber\\
  &=&	\frac{1-(w^{n+1}/c_{s-1\,r})}{1-(w^{n+1}/c_{sr})}\,
    \frac{1-(w^{n+1}/c_{s\,r-1})}{1-(w^{n+1}/c_{s-1\,r-1})},\label{eqYnsr}\\
  &&\hspace{25ex} (1\leq r<s-1<N\,,\, (r,s) \neq (1,N)),\nonumber\\
  y^{(n)}_{r+1\,r}&=&f(x_{r-1}w^n;x_{r+1},y^{(0)}_{r+1\,r-1},0,x_r)
	  \nonumber\\
  &=&	\frac{1-w^{n+1}}{1-(w^{n+1}/x_{r+1})}\,
	\frac{1-(w^{n+1}/x_rx_{r+1})}{1-(w^{n+1}/x_r)}\,,
	\quad (0\leq r<N,\,x_0\equiv x_N),\\
  u^{(n)}_r&=& f(x_{r-1}w^n;x_r,y^{(0)}_{r\,r-1},0,x_r)\nonumber\\
  &=&	\frac{(1-x_rw^{n+1})(1-(w^{n+1}/x_r))}{(1-w^{n+1})^2}\,,
	\quad (1\leq r \leq N).\label{eqU}
\end{eqnarray}

\begin{figure}
\epsfxsize140mm
\centerline{\epsfbox{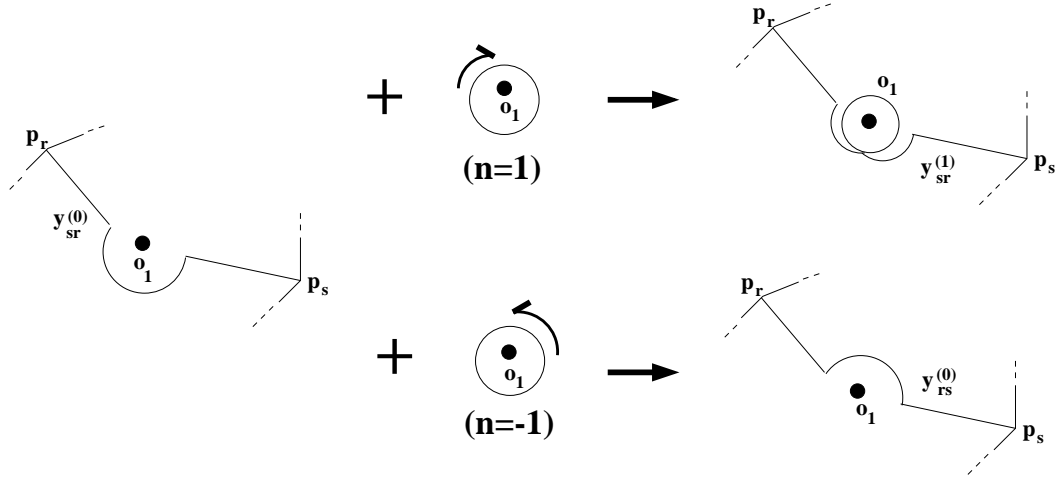}}
\caption{Relations between lines in (dual) diagram.}
\label{figYtoY}
\end{figure}

	We have seen that non-winding lines connecting a pair of
vertices are accompanied with winding lines and this is also the case
for the sides of the $N$-polygon. In Figure \ref{figWL}(b) we see that
$y^{(1)}_{rr+1}$ is another topologically inequivalent line to those
we have seen so far, and the corresponding non-winding line is
a side of the $N$-polygon, $y^{(0)}_{rr+1} (=0)$.
	We can obtain $y^{(n)}_{rr+1}\, (n=1,2,\cdots)$ as follows.
Figure \ref{figYtoY} indicates that $y^{(0)}_{sr}$ with a clockwise
winding gives $y^{(1)}_{sr}$ while $y^{(0)}_{sr}$ with a
counterclockwise winding gives $y^{(0)}_{rs}$, which implies
\begin{equation}
	y^{(-1)}_{sr} = y^{(0)}_{rs}\,.\label{eqYtoY0}
\end{equation}
Assuming that eq.(\ref{eqYnsr}) holds for $n<0$ we can easily prove
eq.(\ref{eqYtoY0}) or we obtain,
\begin{equation}
	y^{(-1-n)}_{sr} = y^{(n)}_{rs}\,.\label{eqYtoY}
\end{equation}
Then we get ($n=1,2,\cdots$),
\begin{eqnarray}
  y^{(n)}_{rr+1}&=&  y^{(-1-n)}_{r+1r} = \frac{1-w^{-n}}{1-(w^{-n}/x_{r+1})}\,
	\frac{1-(w^{-n}/x_rx_{r+1})}{1-(w^{-n}/x_r)}\nonumber\\
  &=&	\frac{1-w^n}{1-x_{r+1}w^n}\,
	\frac{1-x_rx_{r+1}w^n}{1-x_rw^n}\,,
	\quad (0\leq r<N,\,x_0\equiv x_N)\,.\label{eqYrr1}
\end{eqnarray}

\begin{figure}
\centerline{\epsfbox{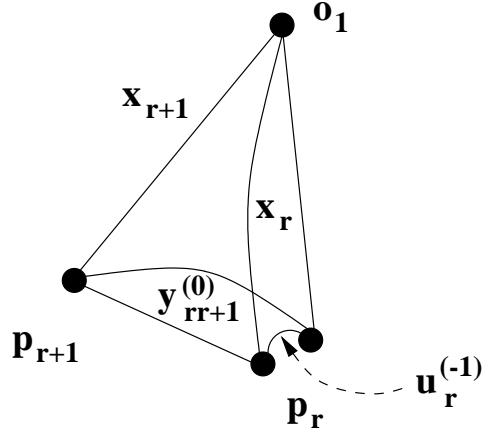}}
\caption{A ``quadrilateral''. $x_r, y^{(0)}_{rr+1}$ and $p_r$ are
written double to be explained.}
\label{figYrr1}
\end{figure}

Three comments are in order:
i) (\ref{eqU}) satisfies
\begin{equation}
	u^{(-2-n)}_r = u^{(n)}_r\,, \quad(n=0,1,2,\cdots).
\end{equation}
ii) Plugging $n=0$ in (\ref{eqYrr1}) we obtain $y^{(0)}_{rr+1}=0$.
iii) If we regard $y^{(0)}_{rr+1}$ as a diagonal of a ``quadrilateral''
$p_rp_rp_{r+1}o_1$ whose sides are ($x_r,u^{(-1)}_r,0,x_{r+1}$) and
the other diagonal is $x_r$ (Figure 13), we have,
\begin{equation}
	y^{(0)}_{rr+1}=f(x_r; x_r,u^{(-1)}_r,0,x_{r+1})\,,
	\label{eqYrr1a}
\end{equation}
or
\begin{eqnarray}
  y^{(0)}_{rr+1}&=& \frac{1-x_r\alpha}{1-x_r\alpha x_r}\,
	\frac{1-x_r\alpha x_r x_{r+1}}{1-x_r\alpha x_{r+1}}\,,
	\label{eqYrr1b}\\
  u^{(-1)}_r &=& \frac{(1-\alpha)(1-\alpha x_r^2)}{(1-\alpha
	x_r)^2}\,. \label{eqUm1}
\end{eqnarray}
  From eq.(\ref{eqU}) with $n=-1$ and eq.(\ref{eqUm1}) we have
$\alpha=1/x_r$, and hence (\ref{eqYrr1b}) becomes $y^{(0)}_{rr+1}=0$.
Then eq.(\ref{eqYrr1a}) leads to ($n=1,2,\cdots$)
\begin{eqnarray}
  y^{(n)}_{rr+1}&=&f(x_rw^n; x_r,u^{(-1)}_r,0,x_{r+1})
	\nonumber\\
  &=& \frac{1-x_rw^n(1/x_r)}{1-x_rw^n(1/x_r) x_r}\,
	\frac{1-x_rw^n(1/x_r) x_r x_{r+1}}{1-x_rw^n(1/x_r) x_{r+1}}
	\nonumber\\
  &=& \frac{1-w^n}{1-x_rw^n}\,
	\frac{1-x_rx_{r+1}w^n}{1-x_{r+1}w^n}\,,
\end{eqnarray}
which agrees with (\ref{eqYrr1}).

	Since we have obtained all the topologically inequivalent
lines, we can now write down the $N$-point one-loop amplitude by the
KSV method. The square of the momentum associated with a line depends
only on the ends of the line and hence, for example, the exponent for
$y^{(n)}_{rs}$ is the same as that for $y^{(n)}_{sr}$ and it is given
by $-\alpha(-(p_r-p_s)^2)-1$, where $\alpha$ is given by
eq.(\ref{3.7}).   The $N$-point one-loop amplitude is given by,
\begin{eqnarray}
  A_N^{(1)}&=& g^N \int d^{26}k \int_{0}^{1}
	\prod_{i=1}^{N} dx_i\, 	\rho^{(1)}(x_i)\,
	 \prod_{i=1}^{N}\, x_i^{-\alpha(-(k-p_i)^2)-1}\,
	\prod_{1\leq r < s-1 < N \atop (r,s)\neq(1,N)}
	 \left( \prod_{n=0}^{\infty}y^{(n)}_{rs}\,
	y^{(n)}_{sr}\right)^{-\alpha(-(p_r-p_s)^2)-1}\nonumber \\
  && \times \prod_{r=0}^{N-1}
	\left(\prod_{n=0}^{\infty}y^{(n)}_{r+1r}\,
	\prod_{n=1}^{\infty}y^{(n)}_{rr+1}\right)
	^{-\alpha(-(p_r-p_{r+1})^2)-1}\,
	\prod_{r=1}^{N}\left(\prod_{n=0}^{\infty} u^{(n)}_r \right)
	^{-\alpha (0) -1}\,,
\end{eqnarray}
where $\rho^{(1)}(x_i)$ is given by,
\begin{equation}
	\rho^{(1)}(x_i)\propto \prod_{1\leq r<s\leq N} (1 - x_{r-1}x_r)^{-1}
	\,, \quad(x_0\equiv x_N),
\end{equation}
due to crossing symmetry.
The amplitude in the operator formalism is given by (see e.g.\cite{GSW}),
\begin{eqnarray}
  A_{Nop}^{(1)}&=& g^N \int d^{26}p \int_{0}^{1}
    \prod_{i=1}^{N} dX_i\, \left(\frac{1}{W^2}\right)\,
	\prod_{i=1}^{N}\,X_i^{\alpha'q_i^2}\,
	\prod_{n=1}^{\infty}(1 - W^n)^{-24}\nonumber\\
  &&\times  \prod_{1\leq r<s\leq N}
    \left\{ \prod_{n=1}^{\infty}\exp\left(-\frac{C_{sr}^n +
    (W/C_{sr})^n - 2W^n}{n (1-W^n)}\right)\right\}^{P_r\cdot P_s}\nonumber \\
  &=& g^N \int d^{26}p \int_{0}^{1} \prod_{i=1}^{N} dX_i\,
	\prod_{i=1}^{N}\, X_i^{-\alpha(-(p+p_N-p_{i-1})^2)-1}\,
	\prod_{n=1}^{\infty}(1 - W^n)^{-24}\nonumber\\
  &&\times  \prod_{1\leq r<s\leq N}
    \left\{(1-C_{sr}) \prod_{n=1}^{\infty}\,
    \left( \frac{(1-C_{sr}W^n)(1-W^n/C_{sr})}{(1-W^n)^2}
	\right)\right\}^{P_r\cdot P_s},
\end{eqnarray}
where
\begin{eqnarray}
	q_r&=&p-P_1 - P_2 - \cdots - P_{r-1}=p+p_N - p_{r-1},\\
	C_{sr}&=& X_{r+1} X_{r+2} \cdots X_{s-1} X_s,\\
	W&=& X_1 X_2 \cdots X_N.
\end{eqnarray}
Considering the correspondence of the parameters,
\begin{eqnarray*}
	k&\longleftrightarrow& p+p_N,\\
	x_i&\longleftrightarrow&X_{i+1},\\
        c_{sr}&\longleftrightarrow&C_{s+1r+1},\\
        w&\longleftrightarrow&W,
\end{eqnarray*}
we shall compare the amplitudes. Then we find,
\begin{equation}
	\rho^{(1)}(x_i)= \left[\prod_{n=1}^{\infty}
	 (1 - w^n) \right]^{-24}\,
	\prod_{1\leq r<s\leq N} (1 - x_{r-1}x_r)^{-1}.
\end{equation}

%%%%%%%%%%%%%%%%%%%%%%%%%%%%%%%%%%%%%%%%%%%%%%%%%%%%%%%%%%%%%%

%%%%%%%%%%%%%%%%%%%%%%%%%%%%%%%%%%%%%%%%%%%%%%%%%%%%%%%%%%%%%%%%%%%%
%%%%%%%%%%%%%%%%%%%%%%%%%%%%%%%%%%%%%%%%%%%%%%%%%%%%%%%%%%%%%%%%%%%%
\end{document}